\documentclass[journal=jpccck,manuscript=article,layout=twocolumn]{achemso}

\usepackage[utf8]{inputenc}
\usepackage[T1]{fontenc} 
\usepackage{amsmath}
\usepackage{amssymb}

\begin{tocentry}
	\includegraphics[width=8.25 cm]{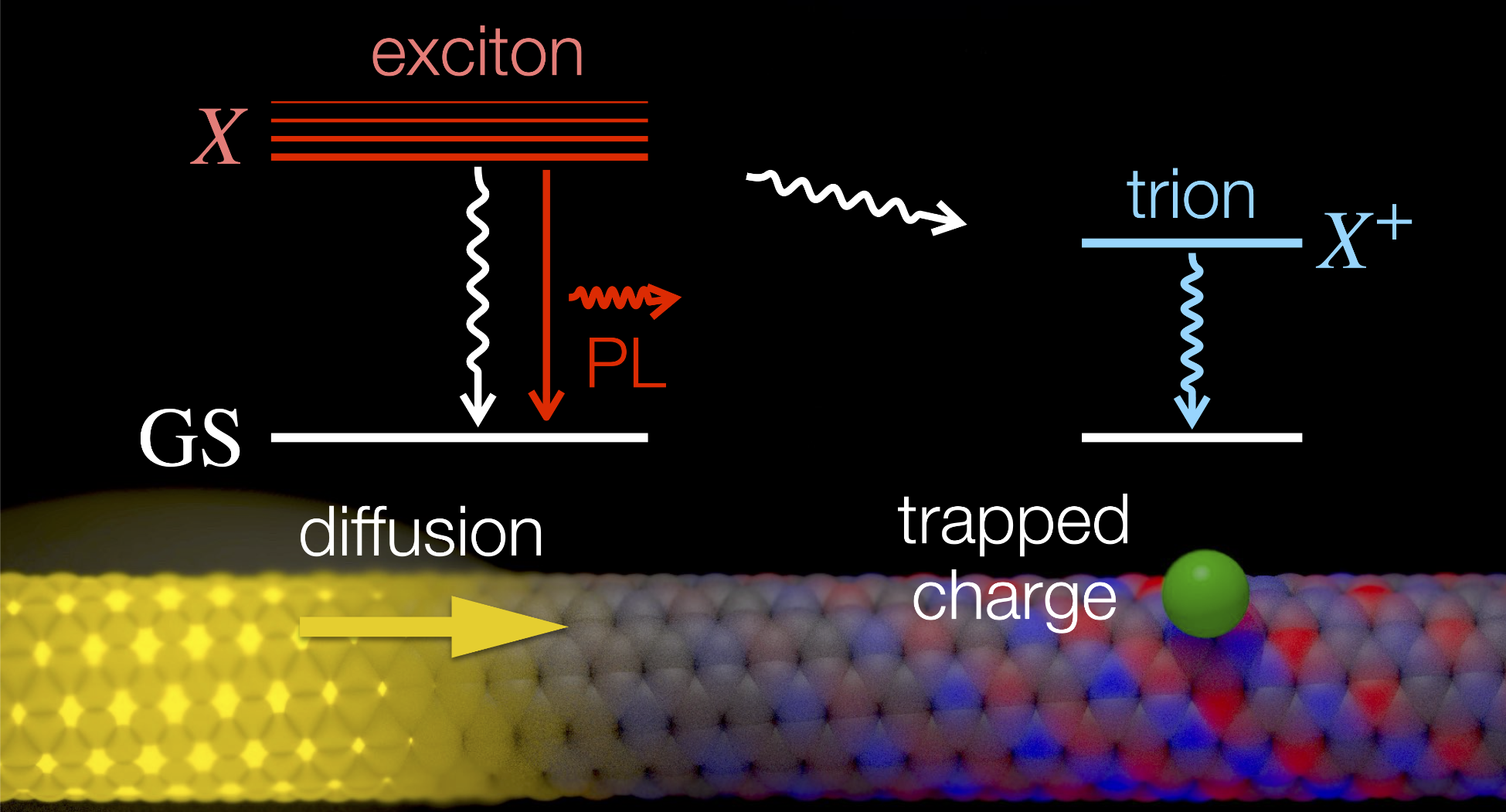}
\end{tocentry}

\author{Klaus H. Eckstein}
\affiliation[uniwue]{Institute of Physical and Theoretical Chemistry, Julius-Maximilian University W\"urzburg, Germany}
\phone{+49 931 3186300}
\author{Pascal Kunkel}
\affiliation[uniwue]{Institute of Physical and Theoretical Chemistry, Julius-Maximilian University W\"urzburg, Germany}
\author{Markus Voelckel}
\affiliation[uniwue]{Institute of Physical and Theoretical Chemistry, Julius-Maximilian University W\"urzburg, Germany}
\author{Friedrich Sch\"oppler}
\affiliation[uniwue]{Institute of Physical and Theoretical Chemistry, Julius-Maximilian University W\"urzburg, Germany}
\author{Tobias Hertel}
\affiliation[uniwue]{Institute of Physical and Theoretical Chemistry, Julius-Maximilian University W\"urzburg, Germany}
\email{tobias.hertel@uni-wuerzburg.de}

\title{Trions, Exciton Dynamics and Spectral Modifications in Doped Carbon Nanotubes: A Singular Defect-Driven Mechanism}

\keywords{doping, defect, trap, carbon nanotube, semiconductor, exciton, trion, spectroscopy }

\begin{document}

\begin{abstract}

Doping substantially influences the electronic and photophysical properties of semiconducting single-wall carbon nanotubes (s-SWNTs). Although prior studies have noted that surplus charge carriers modify optical spectra and accelerate non-radiative exciton decay in doped s-SWNTs, a direct mechanistic correlation of trion formation, exciton dynamics and energetics remains elusive. This work examines the influence of doping-induced non-radiative decay and exciton confinement on s-SWNT photophysics. Using photoluminescence, continuous-wave absorption, and pump-probe spectroscopy, we show that localization of and barrier formation by trapped charges can be jointly quantified using diffusive exciton transport- and particle-in-the-box models, yielding a one-to-one correlation between charge carrier concentrations derived from these models. The study highlights the multifaceted role of exohedral counterions, which trap charges to create quenching sites, form barriers to exciton movement, and host trion states. This contributes significantly to understanding and optimizing the photophysical properties of doped SWNTs.
\end{abstract}

\section{Introduction}

Doped and intrinsic semiconducting single-wall carbon nanotubes (s-SWNTs) are considered as promising materials for electronics \cite{Bishop2020, Koo2017}, sensors \cite{Chen2016}, photonics \cite{Avouris2008, He2018, Ishii2018}, photovoltaic \cite{Ren2011, Kubie2018, Jain2012}, and bioimaging applications \cite{Danne2018, Farrera2017, Pan2017}. Such broad interest can be attributed to favorable charge carrier transport properties, large exciton and trion (charged exciton) oscillator strengths, and near-infrared photoluminescence (PL). 

Yet, photoluminescence quantum yields (PLQYs) on the order of a few percent as well as short excited state lifetimes pose challenges for potential applications, primarily due to the efficiency of non-radiative (NR) decay in s-SWNTs \cite{Wang2004, Carlson2007, Hertel2010, Crochet2007}. This motivates research into the origins and detailed mechanisms of NR decay in these systems.

In general, the decay of excited states in extended systems can be categorized as either homogeneous, {\it i.e.}, identical for all excited states irrespective of their location, or inhomogeneous, {\it i.e.}, dependent on the excited state's localization and location. While homogeneous decay occurs uniformly, inhomogeneous decay is attributed to perturbations of the ideal crystalline structure of a system. In the context of carbon nanotubes, this could be due to interactions of excited states with a disordered environment.

In spite of their pronounced environmental sensitivity, the non-radiative decay of excitons and trions in carbon nanotubes has often been characterized using homogeneous models. These involve couplings between bright excitons, dark excitons, trions, and the ground state, based on unimolecular- or, in high excitation density scenarios, bimolecular rate-constants \cite{Wang2004b, Berciaud2008, Koyama2013, Nishihara2013, Akizuki2014}.
 
In contrast, other studies have favored an inhomogeneous description, specifically regarding electronic transport phenomena \cite{Murrey2023} or the mechanism of NR exciton decay \cite{Hertel2010, Eckstein2017, Ishii2019, Hertel2019, Birkmeier2022}. Such models were particularly successful in describing the dependence of fluorescence quantum yields in SWNTs on nanotube length \cite{Miyauchi2010, Hertel2010, Ishii2015, Graf2016, Ishii2019}, the effect of defects on PL-imaging of SWNTs \cite{Crochet2012, Hartmann2015}, and the influence of diffusive exciton transport on NR decay observed in time-domain investigations of exciton dynamics \cite{Russo2006, Langlois2015, Zhu2007, Wang2017, Eckstein2017, Bai2018}. Moreover, localization of excess charge carriers has also been instrumental for the interpretation of Breit-Wigner-Fano resonances in the IR spectra of doped SWNTs \cite{Lapointe2012, Eckstein2021}. 

While past studies have individually highlighted the impact of localized defects on various optical phenomena there has been no concerted effort to fully explore the congruence among these findings. The diffusion and confinement models have yet to be  utilized for a comparison of the role of exohedral counterions in trapping charges. Such trapped charges are believed to act as quenching sites, form barriers to exciton transport, and support trion state formation. Moreover, a comprehensive assessment of charge carrier densities, derived from individual studies, has been overlooked. As a result, a unified view of how various doping-induced optical phenomena interrelate at differing doping levels remains elusive.

In this work, we provide compelling evidence suggesting that the dynamical and spectral changes observed in the optical spectra of doped semiconducting carbon nanotubes are predominantly governed by \textit{inhomogeneous} doping. These changes, we demonstrate, can be quantitatively correlated using diffusive exciton transport and particle-in-a-box confinement models. Building upon previous findings, we recognize that charges can become trapped at defects within s-SWNTs, formed through interactions with exohedral counterions either adsorbed on or residing near nanotube surfaces \cite{Eckstein2017, Eckstein2019, Eckstein2021}. Not only does our study reinforce evidence for exciton scavenging at such defect sites, but it also presents a detailed prediction of the exciton and trion PLQY dependence on doping levels. The key accomplishment of our research lies in demonstrating a quantitative agreement between spectral and dynamical analyses of doping-induced changes, grounded in the application of the aforementioned models. This breakthrough underscores the importance of inhomogeneous decay mechanisms in understanding and optimizing the potential of s-SWNTs across various applications.

\section{Methods}

Nanotube samples were fabricated as described previously~\cite{Eckstein2019} from organic (6,5) s-SWNT enriched suspensions of PFO-BPy polymer-stabilized (American Dye Source) CoMoCAT nanotubes (SWeNT SG 65, Southwest Nano Technologies Inc.). This procedure leads to a (6,5) purity of 93\% based on absorption spectroscopy (see Supporting Information for details) and average tube lengths of $\approx 400\,\rm nm$ as reported in the literature~\cite{Graf2016,Bottacchi2016,Wang2017}. Redox-chemical doping of s-SWNTs in suspension (5:1 toluene to acetonitrile solvent mixture) was achieved by titration of aliquots of gold(III) chloride (Sigma-Aldrich, $\geq$\,99.99\%) solution using the same solvent mixture. The description of the optical setups used for stationary and time-resolved absorption spectroscopy as well as for photoluminescence measurements (PL) can be found elsewhere \cite{Hartleb2015, Eckstein2017}. The detector setup used for near infrared time-resolved spectroscopy of the trion band was described by Shi et al. \cite{Shi2018}. 

The development and editing of this manuscript was supported by the use of OpenAI's GPT-4 and GPT-3.5-turbo large language models (LLMs). These LLMs were interacted with through two interfaces: the ChatGPT browser-based command line interface and the OpenAI Application Programming Interface (API). The goal of these interactions was to elicit assistance from the LLMs in proofreading paragraphs of human-written text. A representative prompt provided to the LLMs was structured as follows: '\textit{You are the editor of a highly esteemed journal in the physical sciences. Proofread the provided text and amend as required to enhance structural coherence, narrative flow, and clarity of argument. Adherence to the American Institute of Physics (AIP) style guidelines is essential, and the language used should be suitable for a reader with advanced understanding in the physical sciences.}' The revised texts were iteratively refined to ensure that the original ideas and intent of its human authors were preserved. 

\section{Results and Discussion}

\begin{figure}[htbp]
	\centering
		\includegraphics[width=8.4 cm]{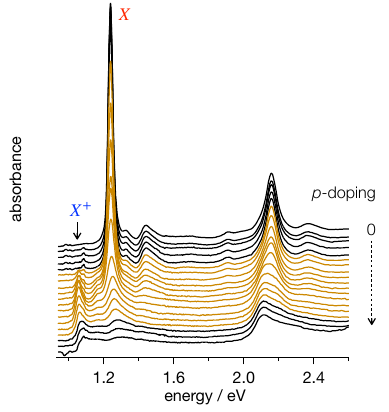}
		\caption{{\bf Absorption spectra of {\it p}-doped SWNTs.}  Waterfall plot of absorption spectra from a suspension of doped (6,5) s-SWNTs, starting with an intrinsic sample on top and ending with a degenerately doped sample at the bottom. $\rm AuCl_3$ concentrations used for doping were 0.00, 0.03, 0.05, 0.07,0.10, \textit{0.30, 0.50, 0.70, 0.90, 1.0, 1.2, 1.5, 2.0, 2.5, 3.0, 4.0,} 6.0, 8.0 and 15 $\rm \mu g/ml$ (top to bottom). Concentrations for spectra highlighted in orange are italicized.}
		\label{fig1}
\end{figure}

\textbf{Absorption spectroscopy.} Figure \ref{fig1} shows a waterfall plot of absorption spectra from a suspension of {\it p}-doped (6,5) s-SWNTs with doping levels increasing from top to bottom. The first subband exciton $X$ at 1.239\,eV loses practically all of its oscillator strength while becoming broader, more asymmetric and shifting to higher energies by about 70\,meV before disappearing into a broad and featureless absorption background at the highest doping levels \cite{Eckstein2017,Bai2018, Eckstein2019}. 

Changes in the second subband range are considerably more subtle. The second exciton subband at 2.158\,eV appears to split into two features, one of which undergoes similar blue-shift as the first subband exciton, while the other feature seems to capture most of the oscillator strength and shifts to slightly lower energies.

\begin{figure}[htbp]
	\centering
		\includegraphics[width=8.4 cm]{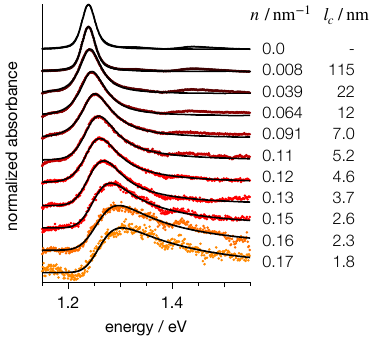}
		\caption{{\bf Quantitative analysis of exciton band position, asymmetry and width.} Normalized exciton absorption spectra are shown for increasing doping levels (top to bottom). The spectra are fit by the confinement model discussed in the main text.}
		\label{fig2}
\end{figure}

The blue shift and increasing asymmetry of the exciton band at higher doping levels have previously been attributed to the increasing confinement of exciton wavefunctions by perturbations introduced through randomly distributed proximal counterions in the environment of the nanotubes \cite{Eckstein2017, Hertel2019}. In Figure \ref{fig2}, we have reproduced a quantitative analysis of normalized absorption spectra using this confinement model, following the discussion by \citeauthor{Hertel2019}\cite{Hertel2019}. The two model parameters used in the fit are the reciprocal concentration of defects $n^{-1}$, which leads to the perturbation of exciton wavefunctions, and their spatial extent in the direction of the nanotube axis $\xi_d$, which is typically found to be near 4\,nm for such samples. The average confinement length $\bar l_c$ responsible for the exciton blue-shift is then given by $\bar l_c=n^{-1}-\xi_d$ (see Fig. \ref{fig2}).

Absorption spectra also feature a doping-induced band at 1.058\,eV, which is frequently attributed to a three-particle trion \cite{Matsunaga2011, Nishihara2012, Park2012, Koyama2013, Mouri2013, Akizuki2014, Eckstein2017, Eckstein2019, Gaulke2020, Eckstein2021} and sometimes to a many-particle exciton-polaron state \cite{Bai2018,Eremin2019}. We designate this band as the $X^+$ trion, in line with the prevailing sentiment in the literature as well as our earlier findings, which do not reveal any noticeable shift of the band position with doping level \cite{Eckstein2017, Eckstein2021}. 

The attribution of the doping-induced band at 1.058\,eV to a trion state is also consistent with the previously reported inhomogeneous nature of redox-chemical doping of (6,5) s-SWNTs, with surplus charges being trapped by shallow Coulomb wells \cite{Eckstein2017, Eckstein2019, Eckstein2021}. These wells support a nearly free-particle-like hole state as the third player for the three-particle trion wave-function \cite{Eckstein2017}. Accordingly, the trion band is generally considered to be a hallmark for the presence of excess charge carriers in doped semiconductors, as is the decrease of exciton oscillator strength \cite{Huard2000, Esser2001}. In the following, we define the doping levels as \textit{moderate} if the trion band has a significant oscillator strength, above about 15\% of its peak intensity and as \textit{low} for smaller doping levels.

\begin{figure}[htbp]
	\centering
		\includegraphics[width=8.4 cm]{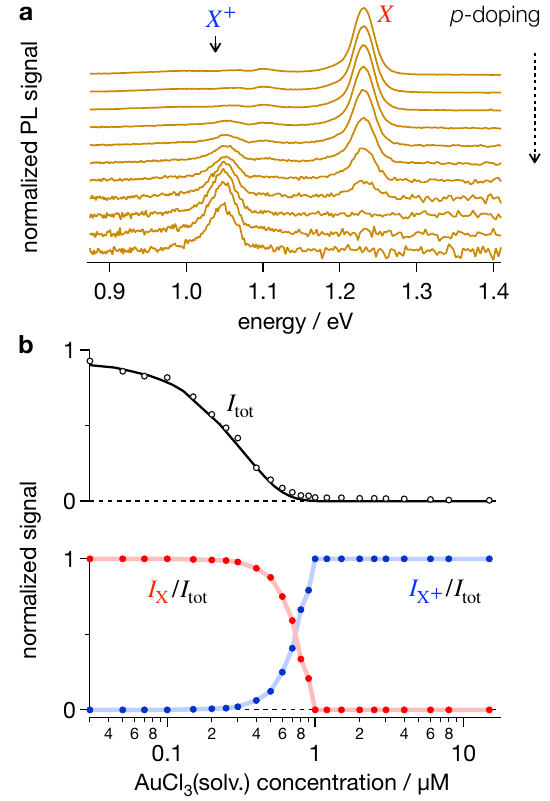}
		\caption{{\bf Photoluminescence from {\it p}-doped SWNTs.} {\bf a)} Waterfall plot with normalized photoluminescence spectra, corresponding to colorized absorption spectra in Fig. \ref{fig1}. {\bf b)} Change of the total PL intensity $I_{\rm PL}$ (top panel) and the normalized exciton and trion signals (bottom panel).}
		\label{fig3}
\end{figure}

\textbf{Photoluminescence Spectra.} The effect of doping on NR exciton and trion decay is seen in the PL spectra shown in Figure \ref{fig3}a) \cite{Hartleb2015, Gaulke2020, Zorn2020}. Here, PL spectra are normalized to the spectrally integrated PL intensity to facilitate a comparison of changes to the relative intensities of emission bands. The decrease of the total photoluminescence intensity with doping level is shown in the upper panel of Figure \ref{fig3}b). The lower panel shows the transfer of relative spectral weight from exciton to trion PL, $I_{\rm X}$ and $I_{\rm X^+}$, respectively, again normalized to the total PL signal $I_{\rm tot}$.

In molecular systems, the concurrent emission of light from different electronic states of a given multiplicity, as illustrated in Figure \ref{fig3}a), is generally prohibited by Kasha's rule. This is mechanistically attributed to the rapid timescale of internal conversion (IC) between excited states, which is faster than the non-radiative decay of the lowest dipole-accessible state. However, in this instance, this does not appear to be the case. 

The PL emission from the exciton, as shown in Figure \ref{fig3}a), is notably more pronounced, particularly at lower doping levels. As we will discuss later, this observation aligns well with the diffusive—or characteristically slow—exciton migration to localized quenching sites, a process that must occur before any energy transfer from the exciton to the trion state can take place. In the inhomogeneously doped system, this energy transfer seems to occur only at specific locations.
\begin{figure}[htbp]
	\centering
		\includegraphics[width=8.4 cm]{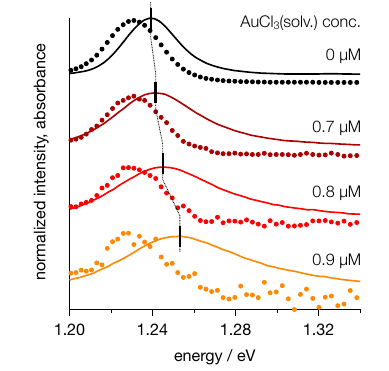}
		\caption{{\bf Exciton absorption band in absorption and emission.} Normalized exciton absorbance (solid lines) and emission spectra (markers) for three different doping levels compared to the intrinsic sample. Concentrations of the solvated dopant are also shown. Absorbance spectra are seen to shift by up to about 12\, meV as indicated by the vertical bars while emission spectra only decrease in intensity.}
		\label{fig4}
\end{figure}

\textbf{Comparison of Absorption and Emission Spectra.} Interestingly, as seen in Figure \ref{fig4}, the exciton PL band shows no significant blue shift in this doping range, even though the absorption band is found to blue-shift by 12\,meV. This observation is difficult to reconcile with a homogeneously doped system, and we interpret it as further evidence of inhomogeneous doping. As discussed above, the blue shift of the absorption band can be explained by the increasing confinement of excitons due to more closely spaced defects \cite{Eckstein2017, Eckstein2019}. Simultaneously, more strongly confined excitons decay more rapidly due to their greater proximity to defects which also act as quenching sites. As a result, the contribution of more strongly confined excitons to the PL spectra decreases. Consequently, PL spectra are dominated by emission from less strongly confined excitons that are barely blue-shifted, require more time to diffuse to a quenching site and thus have a higher probability for radiative decay.
\begin{figure}[htbp]
	\centering
		\includegraphics[width=8.4 cm]{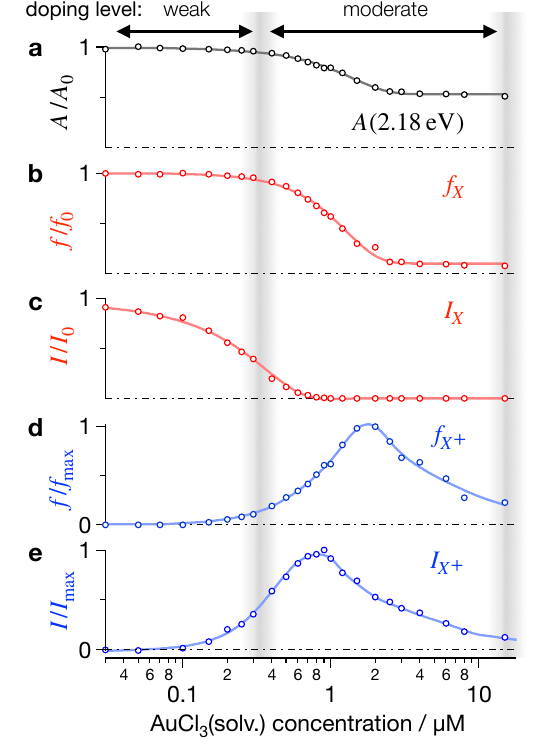}
		\caption{{\bf Overview of doping induced spectral changes.} Doping regimes are indicated at the top. Normalized {\bf a)} absorbance at the second subband exciton energy, {\bf b)} oscillator strength of the first subband exciton, {\bf c)} PL signal intensity of the first subband exciton, {\bf d)} oscillator strength of the trion and {\bf e)} PL signal intensity of the trion.}
		\label{fig5}
\end{figure}

The doping-induced changes of absorption and emission spectra are summarized in Figure \ref{fig5}. It includes normalized data sets for the change of absorbance at the energy of the second subband exciton, the change of first subband exciton and trion oscillator strengths, and the change of their PL signals. Exciton signals are normalized to the oscillator strength and PL intensity of the intrinsic system while trion signals are normalized to the maximum of the respective oscillator strength and PL intensity at moderate doping levels.

The clear correlation between the decrease of the exciton oscillator strength $f_X$ and the rise of trion oscillator strength $f_{X^+}$, as seen in Figure \ref{fig5}b) and d), is noteworthy. This confirms that changes in exciton oscillator strength can justifiably be used as a proxy for assessing doping levels. More specifically, we reported in a previous study that $n=-(0.27\pm 0.02)\, \Delta f/f_0\rm\, nm^{-1}$ represents a good approximation of the doping-induced change in oscillator strength for weak doping levels $n$ \cite{Eckstein2019}.

\begin{figure}[htbp]
	\centering
		\includegraphics[width=8.4 cm]{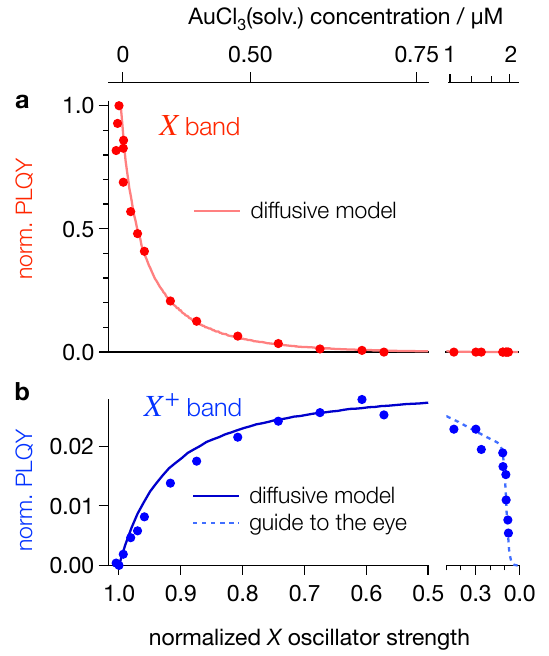}
		\caption{{\bf Dependence of exciton and trion PLQYs on dopant concentration and exciton bleach.} {\bf a)} Normalized change of the first subband exciton PLQY with an inset using an expanded horizontal scale and {\bf b)} relative change of the first subband trion PLQY for excitation at 568\,nm. The blue line is a guide to the eye.}
		\label{fig6}
\end{figure}

The data from Figure \ref{fig5} are next used to investigate the dependence of PLQYs for excitons and trions on the doping level. The normalized exciton PLQY is plotted in Figure \ref{fig6}a) as a function of gold(III) chloride concentration (top axis) and exciton oscillator strength (bottom axis). As seen in Figure \ref{fig6}a), the decrease of PLQY is nearly linear up to a reduction of about 50\%. Therefore, with the above empirical relationship between oscillator strength and doping level, we can also correlate the changes in PLQY $\Delta\Phi/\Phi_0 $ to doping levels using $n=- (0.008\pm 0.002)\cdot \Delta\Phi/\Phi_0 \, \rm nm^{-1}$.

\textbf{Quantitative Analysis of PL Quantum Yields due to NR Exciton Decay.} In the following, the mechanism responsible for the rapid decrease in the PLQY observed in Figure \ref{fig6}a) is investigated. The aim here is to determine if diffusion limited NR decay of excitons at charged defect sites can quantitatively describe the observed decrease. If so, the relationship between PLQYs and doping levels must be modeled according to the rate at which excitons diffuse to charged defects, where non-radiative decay is most likely to occur \cite{Hertel2010, Liu2011, Bai2018}. This diffusion-limited quenching of excitons has previously been used to explain the dependence of PLQYs on nanotube length in intrinsic s-SWNT samples \cite{Hertel2010, Liu2011}. In this context, nanotube ends are considered to act as quenching impurities. However, the influence of doping levels on exciton PLQYs has not been investigated quantitatively using the diffusion-limited quenching model.

Subsequently, PLQYs in doped s-SWNTs are examined using scaling arguments for diffusion-limited quenching at localized quenching sites. Within this framework, the rate of non-radiative decay is determined by the time required for excitons to diffuse to a quenching-defect site. The separation of quenching-defect locations is associated with the inverse of the dopant concentration, $n^{-1}$. 

For correlation with experimental data and specifically with measured changes in oscillator strengths, the relationship between doping levels and oscillator strengths discussed above is used, where $n^{-1}=- \xi_d\,f_0/\Delta f$ and $\xi_d$ is the quenching-defect size. The dependence of exciton PLQYs on the oscillator strength can then be expressed in the following simple form:
\begin{equation}
\frac{\Phi}{\Phi_0}=\left[1+\frac{1-n_0\, \xi_X}{n_0 \left[\xi_d(1+f_0/\Delta f)+\xi_X\right]}\right]^{-2}
\label{eq1}
\end{equation}
where $n_0$ is the concentration of quenching defects for non-doped s-SWNTs (including tube ends), $\xi_X$ is the exciton size, and $\Phi_0$ is the PLQY of the non-doped nanotube (details of the derivation can be found in the supporting information).

The dependence of diffusion time on the square of the diffusion length is reflected by the squared term on the right-hand side of equation \ref{eq1}. To account for the stochastic distribution of quenching-defect locations, the expression in equation \ref{eq1} must be averaged accordingly when fitting experimental data (see supporting information) \cite{Liu2011}.

Best agreement between experimental data in Fig. \ref{fig6}a) and this model is obtained for $n_0 \cdot \xi_d = 0.042 \pm 0.002$, as seen in the red curve in Figure \ref{fig6}a. Using recent estimates of the charged defect wavefunction size in (6,5) SWNTs of $\xi_d = 4\,\rm nm$ \cite{Eckstein2017, Eckstein2019, Eckstein2021}, this result also aligns well with previously reported values of exciton diffusion lengths between 90 and 200\,nm \cite{Cognet2007, Hertel2010, Xie2012, Georgi2009}. The quality of this fit exhibits a comparatively weak dependence on the exciton size, which is here taken to be about $10\,\rm nm$, in accordance with experimental data reported by Mann and Hertel \cite{Mann2016}.

\textbf{Comparison with Trion PL.} As seen in Figure \ref{fig6}b), the PL from the trion state is significantly weaker than that of the exciton, reaching only about 1/40 of the PL intensity originating from intrinsic nanotubes. This implies a very low effective trion PLQY in the range of $10^{-4}-10^{-3}$. However, it's important to note that this value doesn't reflect the intrinsic PLQY of the trion. To understand why, we consider that trions can be generated both directly through light absorption and indirectly via exciton scavenging, with the latter likely to skew measured PLQYs away from the intrinsic value.

As discussed earlier, we found that the decrease in exciton PL can be accurately modeled by diffusive exciton transport to doping-induced quenching sites where trions also form. This non-radiative exciton decay, facilitated by coupling to trion states, is corroborated by the good fit of the trion PL to the diffusive model shown in Fig. \ref{fig6}b, using the same parameters as for the description of the quenching of excitons. At higher doping levels, however, the diffusive model no longer holds, as evidenced by the sudden decrease in trion PL when approaching the degenerate doping regime. The excellent correlation between trion and exciton PL intensities, as a function of doping level, compellingly confirms that localized charges support trion state formation and serve as exciton scavengers.
\begin{figure}[htbp]
	\centering
		\includegraphics[width=8.4 cm]{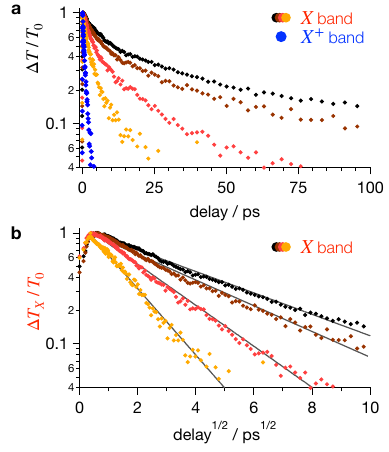}
		\caption{{\bf Exciton dynamics.} {\bf a)} Semi-log plot of the normalized exciton photobleach at $E\rm_{probe}= 1.235\,eV$ in intrinsic (solid black markers) and doped samples (brown to orange) as well as of the trion state in a moderately doped sample (blue markers).   {\bf b)} Exciton photobleach plotted as a function of the square root of time.}
		\label{fig7}
\end{figure}

\textbf{Quantitative Analysis of Exciton Dynamics.} Further examination of non-radiative exciton and trion decay is conducted through femtosecond time-resolved pump-probe experiments, providing direct information on the excited state dynamics. The exciton dynamics, shown in Figures \ref{fig7}a and \ref{fig7}b, were recorded following excitation at 576\,nm. These experiments were carried out at low excitation pulse fluences, in the range of $1 - 8\,\rm \mu J\, cm^{-2}$, corresponding to an exciton density of approximately $(650\,\rm nm)^{-1}$ to limit the impact of exciton-exciton annihilation on the dynamics \cite{Ma2005, Ueda2008, Luer2009}. 

Figure \ref{fig7}a displays pump-probe traces from excitons in both intrinsic and {\it p}-doped nanotubes at varying doping levels up to 100\,ps time-delay, as well as the trion decay in a moderately doped sample. The non-linear decay of the exciton bleach signals in this semi-log plot indicates that the kinetics cannot be described by a unimolecular mechanism, in contrast to predictions from band-filling models of doped SWNTs \cite{Kinder2008,Perebeinos2008,Sau2013}. Instead, the linear decay of the exciton bleach, when plotted as a function of the square root of the time-delay, confirms that the kinetics are diffusion-limited, following a $\exp(-\sqrt{t/\tau_D})$ decay law \cite{Hertel2010, Wang2017, Bai2018}.

Here, the diffusion time, $\tau_D = \pi/(16Dn^2)$, depends on both the diffusion coefficient $D=2.2\,\rm cm^2\, s^{-1}$ and the concentration $n$ of randomly distributed quenching sites \cite{Berezhkovskii1990}. This diffusion coefficient has been obtained from a previous analysis of PLQYs in length selected nanotube samples \cite{Hertel2010} and accounts for the distribution of quenching site separations as discussed by Liu et al.\cite{Liu2011}.

The characteristic diffusion times for the data in Figure \ref{fig7}b are $(20.4\pm 0.4)$\,ps for the intrinsic and $(14.1\pm 0.3)$, $(5.5\pm 0.1)$, and $(1.9\pm 0.1)$\,ps for the increasing doping levels, respectively. These correspond to mean diffusion lengths of 150\, nm in the intrinsic system and 125, 78, and 46\, nm in the progressively more strongly doped nanotubes. The decreasing diffusion times clearly demonstrate how the kinetics remain diffusive but accelerate as the spacing between quenching sites decreases, a finding previously also reported by Bai et al. \cite{Bai2018}.
\begin{figure}[htbp]
	\centering
		\includegraphics[width=8.4 cm]{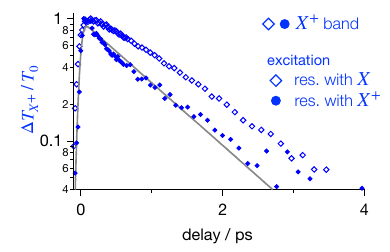}
		\caption{{\bf Trion dynamics.} Semi-log plot of the trion bleach for excitation in resonance with the exciton band and for excitation at the trion resonance.}
		\label{fig8}
\end{figure}

\textbf{Contrasting Observations for Trion Dynamics.} The trion bleach data, which can be seen in Figure \ref{fig8}, were captured for excitation conditions resonant with the exciton at $1000\,\rm nm$ (open markers, excitation density $(70\,\rm nm)^{-1}$, $E\rm_{probe}= 1.065\,eV$) and resonant with the trion band at $1170\,\rm nm$ (solid markers, excitation density $(270\,\rm nm)^{-1}$, $E\rm_{probe}= 1.051\,eV$). The pump-probe trace for resonant excitation of the trion state exhibits unimolecular kinetics with a characteristic rate constant of $1.3\,\rm ps^{-1}$, which is consistent with previous time-resolved experiments \cite{Koyama2013,Nishihara2013,Akizuki2014}. When the exciton band itself is pumped, the decay shown in Figure \ref{fig8} (open markers) is characterized by a slightly slower, yet still unimolecular decay.

These findings suggest two key implications: Firstly, the excitation at the exciton band energy results in some indirect population of the trion state, most likely through the previously mentioned exciton scavenging mechanism. Secondly, the trion population does not seem to be subject to a slow recovery of the ground state, at least not to the extent that this affects the photo-bleach of the exciton band in Figure \ref{fig7}a). This phenomenon is consistent with the previous assertion of partially decoupled ground states of the exciton and trion manifolds \cite{Eckstein2017}, and serves as additional evidence supporting the localized nature of surplus charges at low doping levels \cite{Eremin2019}.

\textbf{Correlation of Carrier Concentrations from Dynamical and Spectral Analyses.} One of the significant findings of this work lies in establishing a direct correlation between dynamical and spectral modifications induced by doping. While both types of modifications have previously been clearly attributed to doping, it has remained unclear whether the observed spectral and dynamical changes can be mechanistically and quantitatively tied to the same or to different microscopic modifications.
\begin{figure}[htbp]
	\centering
		\includegraphics[width=8.4 cm]{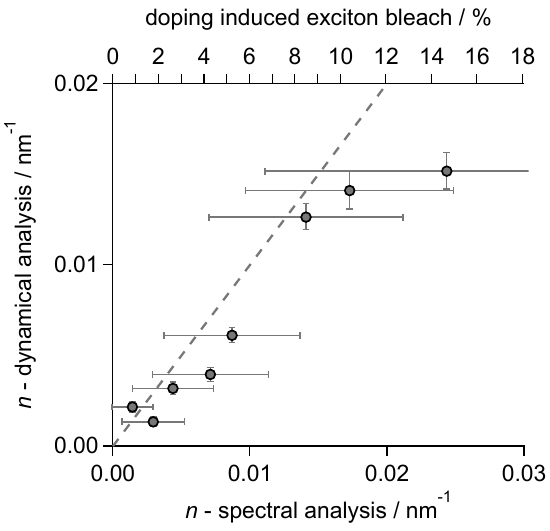}
		\caption{{\bf Correlation of dynamical and spectral analysis}. The vertical axis features doping levels derived from the dynamical analysis of pump-probe data using the diffusion-limited quenching model while the horizontal axis features doping levels derived from the spectral analysis of changes of the exciton band using a particle-in-the-box model. The linear correlation up to concentrations of about $0.015 \rm\, nm^{-1}$ suggests that defects causing exciton quenching and confinement are the very same.}
		\label{fig9}
\end{figure}

To investigate the relationship between these spectral and dynamical changes, we finally use the quenching defect concentrations $n$ as derived from the reciprocal diffusion lengths obtained by the diffusive quenching model, based on the analysis of the pump-probe data discussed above (dynamical analysis). We combine these concentrations for comparison with the defect concentrations $n$ derived from the initially discussed confinement model (spectral analysis) in Fig. \ref{fig9}. This comparison reveals a robust linear one-to-one correlation between quenching site- and confinement site concentrations up to carrier densities of at least $0.015,\rm nm^{-1}$. This leads us to conclude that the defects responsible for exciton confinement and those serving as quenching sites are identical.

For carrier concentrations exceeding those shown in Fig. \ref{fig9}, we anticipate that the dynamical analysis may underestimate the actual concentrations. This expectation is based on the premise that diffusive kinetics provide the most accurate model for exciton transport when the distance between quenching sites is significantly larger than the exciton coherence lengths. The magnitude of these coherence lengths is set by the reported electron-hole correlation lengths, which are on the order of 10\, nm \cite{Mann2016}. As the length scales for diffusive exciton transport approach the spacing between defects, especially at higher carrier concentrations, we expect that the interpretation of pump-probe traces becomes increasingly challenging.

\section{Summary and Conclusions}

Our study provides significant insights into the decay of excitons and trions in doped (6,5) carbon nanotubes, with a particular emphasis on the dominant role of non-radiative quenching at charged defect sites. Notably, a quantitative comparison of the analysis of dynamical and spectral data reveals that defects responsible for exciton quenching are the same as those associated with exciton confinement. Taking into consideration previous research that reported the co-localization of charged defects with optically excited trions \cite{Mouri2013}, this conclusively shows that excitons are quenched by localized charges where trion states are also formed. 

The comprehensive examination of the dependence of exciton and trion photoluminescence quantum yields (PLQYs) and exciton confinement on doping levels, in conjunction with pump-probe spectroscopy of excited state dynamics, lends further support to a unified diffusion-limited non-radiative decay mechanism governing exciton lifetimes. This decay mechanism implies that the rate of diffusive transport of excitons to charged defect sites primarily dictates the energy dissipation rate. At the same time, the results reveal that localized trions are overwhelmingly populated indirectly through exciton scavenging. 

In summary, this research not only enriches our understanding of SWNTs' excited state dynamics but also lays a foundation for the exploration and optimization of these materials for various technological applications, particularly through better controlling Coulomb interactions between surplus charges on the nanotubes and charges in the environment.

\begin{suppinfo}
The Supporting Information is available free of charge at

Derivation of equation (1) describing diffusion-limited PL quenching, sample characterization, and additional transient absorption data.
\end{suppinfo}


\section{Acknowledgements}
K. E. and T.H. acknowledge financial support by the German National Science Foundation through the DFG GRK2112 and through grant HE 3355/4-1.

\bibliography{ms}

\providecommand{\latin}[1]{#1}
\makeatletter
\providecommand{\doi}
  {\begingroup\let\do\@makeother\dospecials
  \catcode`\{=1 \catcode`\}=2 \doi@aux}
\providecommand{\doi@aux}[1]{\endgroup\texttt{#1}}
\makeatother
\providecommand*\mcitethebibliography{\thebibliography}
\csname @ifundefined\endcsname{endmcitethebibliography}
  {\let\endmcitethebibliography\endthebibliography}{}
\begin{mcitethebibliography}{64}
\providecommand*\natexlab[1]{#1}
\providecommand*\mciteSetBstSublistMode[1]{}
\providecommand*\mciteSetBstMaxWidthForm[2]{}
\providecommand*\mciteBstWouldAddEndPuncttrue
  {\def\EndOfBibitem{\unskip.}}
\providecommand*\mciteBstWouldAddEndPunctfalse
  {\let\EndOfBibitem\relax}
\providecommand*\mciteSetBstMidEndSepPunct[3]{}
\providecommand*\mciteSetBstSublistLabelBeginEnd[3]{}
\providecommand*\EndOfBibitem{}
\mciteSetBstSublistMode{f}
\mciteSetBstMaxWidthForm{subitem}{(\alph{mcitesubitemcount})}
\mciteSetBstSublistLabelBeginEnd
  {\mcitemaxwidthsubitemform\space}
  {\relax}
  {\relax}

\bibitem[Bishop \latin{et~al.}(2020)Bishop, Hills, Srimani, Lau, Murphy,
  Fuller, Humes, Ratkovich, Nelson, and Shulaker]{Bishop2020}
Bishop,~M.~D.; Hills,~G.; Srimani,~T.; Lau,~C.; Murphy,~D.; Fuller,~S.;
  Humes,~J.; Ratkovich,~A.; Nelson,~M.; Shulaker,~M.~M. Fabrication of Carbon
  Nanotube Field-Effect Transistors in Commercial Silicon Manufacturing
  Facilities. \emph{Nat. Electron.} \textbf{2020}, \emph{3}, 492--501\relax
\mciteBstWouldAddEndPuncttrue
\mciteSetBstMidEndSepPunct{\mcitedefaultmidpunct}
{\mcitedefaultendpunct}{\mcitedefaultseppunct}\relax
\EndOfBibitem
\bibitem[Koo \latin{et~al.}(2017)Koo, Jeong, Shim, Son, Kim, Kim, Choi, Hong,
  and Kim]{Koo2017}
Koo,~J.~H.; Jeong,~S.; Shim,~H.~J.; Son,~D.; Kim,~J.; Kim,~D.~C.; Choi,~S.;
  Hong,~J.-I.; Kim,~D.-H. Wearable Electrocardiogram Monitor Using Carbon
  Nanotube Electronics and Color-Tunable Organic Light-Emitting Diodes.
  \emph{ACS Nano} \textbf{2017}, \emph{11}, 10032--10041\relax
\mciteBstWouldAddEndPuncttrue
\mciteSetBstMidEndSepPunct{\mcitedefaultmidpunct}
{\mcitedefaultendpunct}{\mcitedefaultseppunct}\relax
\EndOfBibitem
\bibitem[Chen \latin{et~al.}(2016)Chen, Gao, Emaminejad, Kiriya, Ota, Nyein,
  Takei, and Javey]{Chen2016}
Chen,~K.; Gao,~W.; Emaminejad,~S.; Kiriya,~D.; Ota,~H.; Nyein,~H. Y.~Y.;
  Takei,~K.; Javey,~A. Printed Carbon Nanotube Electronics and Sensor Systems.
  \emph{Adv. Mater.} \textbf{2016}, \emph{28}, 4397--4414\relax
\mciteBstWouldAddEndPuncttrue
\mciteSetBstMidEndSepPunct{\mcitedefaultmidpunct}
{\mcitedefaultendpunct}{\mcitedefaultseppunct}\relax
\EndOfBibitem
\bibitem[Avouris \latin{et~al.}(2008)Avouris, Freitag, and
  Perebeinos]{Avouris2008}
Avouris,~P.; Freitag,~M.; Perebeinos,~V. Carbon-Nanotube Photonics and
  Optoelectronics. \emph{Nat. Photonics} \textbf{2008}, \emph{2},
  341--350\relax
\mciteBstWouldAddEndPuncttrue
\mciteSetBstMidEndSepPunct{\mcitedefaultmidpunct}
{\mcitedefaultendpunct}{\mcitedefaultseppunct}\relax
\EndOfBibitem
\bibitem[He \latin{et~al.}(2018)He, Htoon, Doorn, Pernice, Pyatkov, Krupke,
  Jeantet, Chassagneux, and Voisin]{He2018}
He,~X.; Htoon,~H.; Doorn,~S.~K.; Pernice,~W. H.~P.; Pyatkov,~F.; Krupke,~R.;
  Jeantet,~A.; Chassagneux,~Y.; Voisin,~C. Carbon Nanotubes as Emerging
  Quantum-Light Sources. \emph{Nat. Mater.} \textbf{2018}, \emph{17},
  663--670\relax
\mciteBstWouldAddEndPuncttrue
\mciteSetBstMidEndSepPunct{\mcitedefaultmidpunct}
{\mcitedefaultendpunct}{\mcitedefaultseppunct}\relax
\EndOfBibitem
\bibitem[Ishii \latin{et~al.}(2018)Ishii, He, Hartmann, Machiya, Htoon, Doorn,
  and Kato]{Ishii2018}
Ishii,~A.; He,~X.; Hartmann,~N.~F.; Machiya,~H.; Htoon,~H.; Doorn,~S.~K.;
  Kato,~Y.~K. Enhanced Single-Photon Emission from Carbon-Nanotube Dopant
  States Coupled to Silicon Microcavities. \emph{Nano Lett.} \textbf{2018},
  \emph{18}, 3873--3878\relax
\mciteBstWouldAddEndPuncttrue
\mciteSetBstMidEndSepPunct{\mcitedefaultmidpunct}
{\mcitedefaultendpunct}{\mcitedefaultseppunct}\relax
\EndOfBibitem
\bibitem[Ren \latin{et~al.}(2011)Ren, Bernardi, Lunt, Bulovic, Grossman, and
  Grade{\v{c}}ak]{Ren2011}
Ren,~S.; Bernardi,~M.; Lunt,~R.~R.; Bulovic,~V.; Grossman,~J.~C.;
  Grade{\v{c}}ak,~S. Toward Efficient Carbon Nanotube/P3HT Solar Cells: Active
  Layer Morphology, Electrical, and Optical Properties. \emph{Nano Lett.}
  \textbf{2011}, \emph{11}, 5316--5321\relax
\mciteBstWouldAddEndPuncttrue
\mciteSetBstMidEndSepPunct{\mcitedefaultmidpunct}
{\mcitedefaultendpunct}{\mcitedefaultseppunct}\relax
\EndOfBibitem
\bibitem[Kubie \latin{et~al.}(2018)Kubie, Watkins, Ihly, Wladkowski, Blackburn,
  Rice, and Parkinson]{Kubie2018}
Kubie,~L.; Watkins,~K.~J.; Ihly,~R.; Wladkowski,~H.~V.; Blackburn,~J.~L.;
  Rice,~W.~D.; Parkinson,~B.~A. Optically Generated Free-Carrier Collection
  from an All Single-Walled Carbon Nanotube Active Layer. \emph{J. Phys. Chem.
  Lett.} \textbf{2018}, \emph{9}, 4841--4847\relax
\mciteBstWouldAddEndPuncttrue
\mciteSetBstMidEndSepPunct{\mcitedefaultmidpunct}
{\mcitedefaultendpunct}{\mcitedefaultseppunct}\relax
\EndOfBibitem
\bibitem[Jain \latin{et~al.}(2012)Jain, Howden, Tvrdy, Shimizu, Hilmer,
  McNicholas, Gleason, and Strano]{Jain2012}
Jain,~R.~M.; Howden,~R.; Tvrdy,~K.; Shimizu,~S.; Hilmer,~A.~J.;
  McNicholas,~T.~P.; Gleason,~K.~K.; Strano,~M.~S. Polymer-Free Near-Infrared
  Photovoltaics with Single Chirality (6,5) Semiconducting Carbon Nanotube
  Active Layers. \emph{Adv. Mater.} \textbf{2012}, \emph{24}, 4436--4439\relax
\mciteBstWouldAddEndPuncttrue
\mciteSetBstMidEndSepPunct{\mcitedefaultmidpunct}
{\mcitedefaultendpunct}{\mcitedefaultseppunct}\relax
\EndOfBibitem
\bibitem[Dann{\'e} \latin{et~al.}(2018)Dann{\'e}, Godin, Gao, Varela, Groc,
  Lounis, and Cognet]{Danne2018}
Dann{\'e},~N.; Godin,~A.~G.; Gao,~Z.; Varela,~J.~A.; Groc,~L.; Lounis,~B.;
  Cognet,~L. Comparative Analysis of Photoluminescence and Upconversion
  Emission from Individual Carbon Nanotubes for Bioimaging Applications.
  \emph{ACS Photonics} \textbf{2018}, \emph{5}, 359--364\relax
\mciteBstWouldAddEndPuncttrue
\mciteSetBstMidEndSepPunct{\mcitedefaultmidpunct}
{\mcitedefaultendpunct}{\mcitedefaultseppunct}\relax
\EndOfBibitem
\bibitem[Farrera \latin{et~al.}(2017)Farrera, {Torres And{\'o}n}, and
  Feliu]{Farrera2017}
Farrera,~C.; {Torres And{\'o}n},~F.; Feliu,~N. Carbon Nanotubes as Optical
  Sensors in Biomedicine. \emph{ACS Nano} \textbf{2017}, \emph{11},
  10637--10643\relax
\mciteBstWouldAddEndPuncttrue
\mciteSetBstMidEndSepPunct{\mcitedefaultmidpunct}
{\mcitedefaultendpunct}{\mcitedefaultseppunct}\relax
\EndOfBibitem
\bibitem[Pan \latin{et~al.}(2017)Pan, Li, and Choi]{Pan2017}
Pan,~J.; Li,~F.; Choi,~J.~H. Single-Walled Carbon Nanotubes as Optical Probes
  for Bio-Sensing and Imaging. \emph{J. Mater. Chem. B} \textbf{2017},
  \emph{5}, 6511--6522\relax
\mciteBstWouldAddEndPuncttrue
\mciteSetBstMidEndSepPunct{\mcitedefaultmidpunct}
{\mcitedefaultendpunct}{\mcitedefaultseppunct}\relax
\EndOfBibitem
\bibitem[Wang \latin{et~al.}(2004)Wang, Dukovic, Brus, and Heinz]{Wang2004}
Wang,~F.; Dukovic,~G.; Brus,~L.~E.; Heinz,~T.~F. Time-Resolved Fluorescence of
  Carbon Nanotubes and its Implication for Radiative Lifetimes. \emph{Phys.
  Rev. Lett.} \textbf{2004}, \emph{92}, 177401\relax
\mciteBstWouldAddEndPuncttrue
\mciteSetBstMidEndSepPunct{\mcitedefaultmidpunct}
{\mcitedefaultendpunct}{\mcitedefaultseppunct}\relax
\EndOfBibitem
\bibitem[Carlson \latin{et~al.}(2007)Carlson, Maccagnano, Zheng, Silcox, and
  Krauss]{Carlson2007}
Carlson,~L.~J.; Maccagnano,~S.~E.; Zheng,~M.; Silcox,~J.; Krauss,~T.~D.
  Fluorescence Efficiency of Individual Carbon Nanotubes. \emph{Nano Lett.}
  \textbf{2007}, \emph{7}, 3698--3703\relax
\mciteBstWouldAddEndPuncttrue
\mciteSetBstMidEndSepPunct{\mcitedefaultmidpunct}
{\mcitedefaultendpunct}{\mcitedefaultseppunct}\relax
\EndOfBibitem
\bibitem[Hertel \latin{et~al.}(2010)Hertel, Himmelein, Ackermann, Stich, and
  Crochet]{Hertel2010}
Hertel,~T.; Himmelein,~S.; Ackermann,~T.; Stich,~D.; Crochet,~J. Diffusion
  Limited Photoluminescence Quantum Yields in 1-D Semiconductors: Single-Wall
  Carbon Nanotubes. \emph{ACS Nano} \textbf{2010}, \emph{4}, 7161--7168\relax
\mciteBstWouldAddEndPuncttrue
\mciteSetBstMidEndSepPunct{\mcitedefaultmidpunct}
{\mcitedefaultendpunct}{\mcitedefaultseppunct}\relax
\EndOfBibitem
\bibitem[Crochet \latin{et~al.}(2007)Crochet, Clemens, and Hertel]{Crochet2007}
Crochet,~J.; Clemens,~M.; Hertel,~T. Quantum Yield Heterogeneities of Aqueous
  Single-Wall Carbon Nanotube Suspensions. \emph{J. Am. Chem. Soc.}
  \textbf{2007}, \emph{129}, 8058--8059\relax
\mciteBstWouldAddEndPuncttrue
\mciteSetBstMidEndSepPunct{\mcitedefaultmidpunct}
{\mcitedefaultendpunct}{\mcitedefaultseppunct}\relax
\EndOfBibitem
\bibitem[Wang \latin{et~al.}(2004)Wang, Dukovic, Knoesel, Brus, and
  Heinz]{Wang2004b}
Wang,~F.; Dukovic,~G.; Knoesel,~E.; Brus,~L.~E.; Heinz,~T.~F. Observation of
  Rapid Auger Recombination in Optically Excited Semiconducting Carbon
  Nanotubes. \emph{Phys. Rev. B} \textbf{2004}, \emph{70}, 241403(R)\relax
\mciteBstWouldAddEndPuncttrue
\mciteSetBstMidEndSepPunct{\mcitedefaultmidpunct}
{\mcitedefaultendpunct}{\mcitedefaultseppunct}\relax
\EndOfBibitem
\bibitem[Berciaud \latin{et~al.}(2008)Berciaud, Cognet, and
  Lounis]{Berciaud2008}
Berciaud,~S.; Cognet,~L.; Lounis,~B. Luminescence Decay and the Absorption
  Cross Section of Individual Single-Walled Carbon Nanotubes. \emph{Phys. Rev.
  Lett.} \textbf{2008}, \emph{101}, 077402\relax
\mciteBstWouldAddEndPuncttrue
\mciteSetBstMidEndSepPunct{\mcitedefaultmidpunct}
{\mcitedefaultendpunct}{\mcitedefaultseppunct}\relax
\EndOfBibitem
\bibitem[Koyama \latin{et~al.}(2013)Koyama, Shimizu, Miyata, Shinohara, and
  Nakamura]{Koyama2013}
Koyama,~T.; Shimizu,~S.; Miyata,~Y.; Shinohara,~H.; Nakamura,~A. Ultrafast
  Formation and Decay Dynamics of Trions in \textit{p}-Doped Single-Walled
  Carbon Nanotubes. \emph{Phys. Rev. B} \textbf{2013}, \emph{87}, 165430\relax
\mciteBstWouldAddEndPuncttrue
\mciteSetBstMidEndSepPunct{\mcitedefaultmidpunct}
{\mcitedefaultendpunct}{\mcitedefaultseppunct}\relax
\EndOfBibitem
\bibitem[Nishihara \latin{et~al.}(2013)Nishihara, Yamada, Okano, and
  Kanemitsu]{Nishihara2013}
Nishihara,~T.; Yamada,~Y.; Okano,~M.; Kanemitsu,~Y. Trion Formation and
  Recombination Dynamics in Hole-Doped Single-Walled Carbon Nanotubes.
  \emph{Appl. Phys. Lett.} \textbf{2013}, \emph{103}, 023101\relax
\mciteBstWouldAddEndPuncttrue
\mciteSetBstMidEndSepPunct{\mcitedefaultmidpunct}
{\mcitedefaultendpunct}{\mcitedefaultseppunct}\relax
\EndOfBibitem
\bibitem[Akizuki \latin{et~al.}(2014)Akizuki, Iwamura, Mouri, Miyauchi,
  Kawasaki, Watanabe, Suemoto, Watanabe, Asano, and Matsuda]{Akizuki2014}
Akizuki,~N.; Iwamura,~M.; Mouri,~S.; Miyauchi,~Y.; Kawasaki,~T.; Watanabe,~H.;
  Suemoto,~T.; Watanabe,~K.; Asano,~K.; Matsuda,~K. Nonlinear Photoluminescence
  Properties of Trions in Hole-Doped Single-Walled Carbon Nanotubes.
  \emph{Phys. Rev. B} \textbf{2014}, \emph{89}, 195432\relax
\mciteBstWouldAddEndPuncttrue
\mciteSetBstMidEndSepPunct{\mcitedefaultmidpunct}
{\mcitedefaultendpunct}{\mcitedefaultseppunct}\relax
\EndOfBibitem
\bibitem[Murrey \latin{et~al.}(2023)Murrey, Aubry, Ruiz, Thurman, Eckstein,
  Doud, Stauber, Spokoyny, Schwartz, Hertel, Blackburn, and
  Ferguson]{Murrey2023}
Murrey,~T.~L.; Aubry,~T.~J.; Ruiz,~O.~L.; Thurman,~K.~A.; Eckstein,~K.~H.;
  Doud,~E.~A.; Stauber,~J.~M.; Spokoyny,~A.~M.; Schwartz,~B.~J.; Hertel,~T.
  \latin{et~al.}  Tuning Counterion Chemistry to Reduce Carrier Localization in
  Doped Semiconducting Carbon Nanotube Networks. \emph{Cell Rep. Phys. Sci.}
  \textbf{2023}, \emph{4}, 101407\relax
\mciteBstWouldAddEndPuncttrue
\mciteSetBstMidEndSepPunct{\mcitedefaultmidpunct}
{\mcitedefaultendpunct}{\mcitedefaultseppunct}\relax
\EndOfBibitem
\bibitem[Eckstein \latin{et~al.}(2017)Eckstein, Hartleb, Achsnich,
  Sch{\"o}ppler, and Hertel]{Eckstein2017}
Eckstein,~K.~H.; Hartleb,~H.; Achsnich,~M.~M.; Sch{\"o}ppler,~F.; Hertel,~T.
  Localized Charges Control Exciton Energetics and Energy Dissipation in Doped
  Carbon Nanotubes. \emph{ACS Nano} \textbf{2017}, \emph{10},
  10401--10408\relax
\mciteBstWouldAddEndPuncttrue
\mciteSetBstMidEndSepPunct{\mcitedefaultmidpunct}
{\mcitedefaultendpunct}{\mcitedefaultseppunct}\relax
\EndOfBibitem
\bibitem[Ishii \latin{et~al.}(2019)Ishii, Machiya, and Kato]{Ishii2019}
Ishii,~A.; Machiya,~H.; Kato,~Y.~K. High Efficiency Dark-to-Bright Exciton
  Conversion in Carbon Nanotubes. \emph{Phys. Rev. X} \textbf{2019}, \emph{9},
  041048\relax
\mciteBstWouldAddEndPuncttrue
\mciteSetBstMidEndSepPunct{\mcitedefaultmidpunct}
{\mcitedefaultendpunct}{\mcitedefaultseppunct}\relax
\EndOfBibitem
\bibitem[Hertel(2019)]{Hertel2019}
Hertel,~T. In \emph{Optical Properties of Carbon Nanotubes}; Weisman,~R.~B.,
  Kono,~J., Eds.; World Scientific Series on Carbon Nanoscience, Handbook of
  Carbon Nanomaterials; {World Scientific}: New Jersey, 2019; Vol.~10; pp
  191--236\relax
\mciteBstWouldAddEndPuncttrue
\mciteSetBstMidEndSepPunct{\mcitedefaultmidpunct}
{\mcitedefaultendpunct}{\mcitedefaultseppunct}\relax
\EndOfBibitem
\bibitem[Birkmeier \latin{et~al.}(2022)Birkmeier, Hertel, and
  Hartschuh]{Birkmeier2022}
Birkmeier,~K.; Hertel,~T.; Hartschuh,~A. Probing the Ultrafast Dynamics of
  Excitons in Single Semiconducting Carbon Nanotubes. \emph{Nat. Commun.}
  \textbf{2022}, \emph{13}\relax
\mciteBstWouldAddEndPuncttrue
\mciteSetBstMidEndSepPunct{\mcitedefaultmidpunct}
{\mcitedefaultendpunct}{\mcitedefaultseppunct}\relax
\EndOfBibitem
\bibitem[Miyauchi \latin{et~al.}(2010)Miyauchi, Matsuda, Yamamoto, Nakashima,
  and Kanemitsu]{Miyauchi2010}
Miyauchi,~Y.; Matsuda,~K.; Yamamoto,~Y.; Nakashima,~N.; Kanemitsu,~Y.
  Length-Dependent Photoluminescence Lifetimes in Single-Walled Carbon
  Nanotubes. \emph{J. Phys. Chem. C} \textbf{2010}, \emph{114},
  12905--12908\relax
\mciteBstWouldAddEndPuncttrue
\mciteSetBstMidEndSepPunct{\mcitedefaultmidpunct}
{\mcitedefaultendpunct}{\mcitedefaultseppunct}\relax
\EndOfBibitem
\bibitem[Ishii \latin{et~al.}(2015)Ishii, Yoshida, and Kato]{Ishii2015}
Ishii,~A.; Yoshida,~M.; Kato,~Y.~K. Exciton Diffusion, End Quenching, and
  Exciton-Exciton Annihilation in Individual Air-Suspended Carbon Nanotubes.
  \emph{Phys. Rev. B} \textbf{2015}, \emph{91}, 125427\relax
\mciteBstWouldAddEndPuncttrue
\mciteSetBstMidEndSepPunct{\mcitedefaultmidpunct}
{\mcitedefaultendpunct}{\mcitedefaultseppunct}\relax
\EndOfBibitem
\bibitem[Graf \latin{et~al.}(2016)Graf, Zakharko, Schie{\ss}l, Backes, Pfohl,
  Flavel, and Zaumseil]{Graf2016}
Graf,~A.; Zakharko,~Y.; Schie{\ss}l,~S.~P.; Backes,~C.; Pfohl,~M.;
  Flavel,~B.~S.; Zaumseil,~J. Large Scale, Selective Dispersion of Long
  Single-Walled Carbon Nanotubes with High Photoluminescence Quantum Yield by
  Shear Force Mixing. \emph{Carbon} \textbf{2016}, \emph{105}, 593--599\relax
\mciteBstWouldAddEndPuncttrue
\mciteSetBstMidEndSepPunct{\mcitedefaultmidpunct}
{\mcitedefaultendpunct}{\mcitedefaultseppunct}\relax
\EndOfBibitem
\bibitem[Crochet \latin{et~al.}(2012)Crochet, Duque, Werner, Lounis, Cognet,
  and Doorn]{Crochet2012}
Crochet,~J.~J.; Duque,~J.~G.; Werner,~J.~H.; Lounis,~B.; Cognet,~L.;
  Doorn,~S.~K. Disorder Limited Exciton Transport in Colloidal Single-Wall
  Carbon Nanotubes. \emph{Nano Lett.} \textbf{2012}, \emph{12},
  5091--5096\relax
\mciteBstWouldAddEndPuncttrue
\mciteSetBstMidEndSepPunct{\mcitedefaultmidpunct}
{\mcitedefaultendpunct}{\mcitedefaultseppunct}\relax
\EndOfBibitem
\bibitem[Hartmann \latin{et~al.}(2015)Hartmann, Yalcin, Adamska, H{\'a}roz, Ma,
  Tretiak, Htoon, and Doorn]{Hartmann2015}
Hartmann,~N.~F.; Yalcin,~S.~E.; Adamska,~L.; H{\'a}roz,~E.~H.; Ma,~X.;
  Tretiak,~S.; Htoon,~H.; Doorn,~S.~K. Photoluminescence Imaging of Solitary
  Dopant Sites in Covalently Doped Single-Wall Carbon Nanotubes.
  \emph{Nanoscale} \textbf{2015}, \emph{7}, 20521--20530\relax
\mciteBstWouldAddEndPuncttrue
\mciteSetBstMidEndSepPunct{\mcitedefaultmidpunct}
{\mcitedefaultendpunct}{\mcitedefaultseppunct}\relax
\EndOfBibitem
\bibitem[Russo \latin{et~al.}(2006)Russo, Mele, Kane, Rubtsov, Therien, and
  Luzzi]{Russo2006}
Russo,~R.~M.; Mele,~E.~J.; Kane,~C.~L.; Rubtsov,~I.~V.; Therien,~M.~J.;
  Luzzi,~D.~E. One-Dimensional Diffusion-Limited Relaxation of Photoexcitations
  in Suspensions of Single-Walled Carbon Nanotubes. \emph{Phys. Rev. B}
  \textbf{2006}, \emph{74}, 041405\relax
\mciteBstWouldAddEndPuncttrue
\mciteSetBstMidEndSepPunct{\mcitedefaultmidpunct}
{\mcitedefaultendpunct}{\mcitedefaultseppunct}\relax
\EndOfBibitem
\bibitem[Langlois \latin{et~al.}(2015)Langlois, Parret, Vialla, Chassagneux,
  Roussignol, Diederichs, Cassabois, Lauret, and Voisin]{Langlois2015}
Langlois,~B.; Parret,~R.; Vialla,~F.; Chassagneux,~Y.; Roussignol,~P.;
  Diederichs,~C.; Cassabois,~G.; Lauret,~J.-S.; Voisin,~C. Intraband and
  Intersubband Many-Body Effects in the Nonlinear Optical Response of
  Single-Wall Carbon Nanotubes. \emph{Phys. Rev. B} \textbf{2015}, \emph{92},
  155423\relax
\mciteBstWouldAddEndPuncttrue
\mciteSetBstMidEndSepPunct{\mcitedefaultmidpunct}
{\mcitedefaultendpunct}{\mcitedefaultseppunct}\relax
\EndOfBibitem
\bibitem[Zhu \latin{et~al.}(2007)Zhu, Crochet, Arnold, Hersam, Ulbricht,
  Resasco, and Hertel]{Zhu2007}
Zhu,~Z.; Crochet,~J.; Arnold,~M.~S.; Hersam,~M.~C.; Ulbricht,~H.; Resasco,~D.;
  Hertel,~T. Pump-Probe Spectroscopy of Exciton Dynamics in (6,5) Carbon
  Nanotubes. \emph{J. Phys. Chem. C} \textbf{2007}, \emph{111},
  3831--3835\relax
\mciteBstWouldAddEndPuncttrue
\mciteSetBstMidEndSepPunct{\mcitedefaultmidpunct}
{\mcitedefaultendpunct}{\mcitedefaultseppunct}\relax
\EndOfBibitem
\bibitem[Wang \latin{et~al.}(2017)Wang, Shea, Flach, McDonough, Way, Zanni, and
  Arnold]{Wang2017}
Wang,~J.; Shea,~M.~J.; Flach,~J.~T.; McDonough,~T.~J.; Way,~A.~J.;
  Zanni,~M.~T.; Arnold,~M.~S. Role of Defects as Exciton Quenching Sites in
  Carbon Nanotube Photovoltaics. \emph{J. Phys. Chem. C} \textbf{2017},
  \emph{121}, 8310--8318\relax
\mciteBstWouldAddEndPuncttrue
\mciteSetBstMidEndSepPunct{\mcitedefaultmidpunct}
{\mcitedefaultendpunct}{\mcitedefaultseppunct}\relax
\EndOfBibitem
\bibitem[Bai \latin{et~al.}(2018)Bai, Olivier, Bullard, Liu, and
  Therien]{Bai2018}
Bai,~Y.; Olivier,~J.-H.; Bullard,~G.; Liu,~C.; Therien,~M.~J. Dynamics of
  Charged Excitons in Electronically and Morphologically Homogeneous
  Single-Walled Carbon Nanotubes. \emph{Proc. Natl. Acad. Sci. U. S. A.}
  \textbf{2018}, \emph{115}, 674--679\relax
\mciteBstWouldAddEndPuncttrue
\mciteSetBstMidEndSepPunct{\mcitedefaultmidpunct}
{\mcitedefaultendpunct}{\mcitedefaultseppunct}\relax
\EndOfBibitem
\bibitem[Lapointe \latin{et~al.}(2012)Lapointe, Gaufr{\`e}s, Tremblay, Tang,
  Martel, and Desjardins]{Lapointe2012}
Lapointe,~F.; Gaufr{\`e}s,~{\'E}.; Tremblay,~I.; Tang,~N. Y.-W.; Martel,~R.;
  Desjardins,~P. Fano Resonances in the Midinfrared Spectra of Single-Walled
  Carbon Nanotubes. \emph{Phys. Rev. Lett.} \textbf{2012}, \emph{109},
  097402\relax
\mciteBstWouldAddEndPuncttrue
\mciteSetBstMidEndSepPunct{\mcitedefaultmidpunct}
{\mcitedefaultendpunct}{\mcitedefaultseppunct}\relax
\EndOfBibitem
\bibitem[Eckstein \latin{et~al.}(2021)Eckstein, Hirsch, Martel, and
  Hertel]{Eckstein2021}
Eckstein,~K.~H.; Hirsch,~F.; Martel,~R.; Hertel,~T. Infrared Study of Charge
  Carrier Confinement in Doped (6,5) Carbon Nanotubes. \emph{J. Phys. Chem. C}
  \textbf{2021}, 5700--5707\relax
\mciteBstWouldAddEndPuncttrue
\mciteSetBstMidEndSepPunct{\mcitedefaultmidpunct}
{\mcitedefaultendpunct}{\mcitedefaultseppunct}\relax
\EndOfBibitem
\bibitem[Eckstein \latin{et~al.}(2019)Eckstein, Oberndorfer, Achsnich,
  Sch{\"o}ppler, and Hertel]{Eckstein2019}
Eckstein,~K.~H.; Oberndorfer,~F.; Achsnich,~M.~M.; Sch{\"o}ppler,~F.;
  Hertel,~T. Quantifying Doping Levels in Carbon Nanotubes by Optical
  Spectroscopy. \emph{J. Phys. Chem. C} \textbf{2019}, \emph{123},
  30001--30006\relax
\mciteBstWouldAddEndPuncttrue
\mciteSetBstMidEndSepPunct{\mcitedefaultmidpunct}
{\mcitedefaultendpunct}{\mcitedefaultseppunct}\relax
\EndOfBibitem
\bibitem[Bottacchi \latin{et~al.}(2016)Bottacchi, Bottacchi, Sp{\"a}th, Namal,
  Hertel, and Anthopoulos]{Bottacchi2016}
Bottacchi,~F.; Bottacchi,~S.; Sp{\"a}th,~F.; Namal,~I.; Hertel,~T.;
  Anthopoulos,~T.~D. Nanoscale Charge Percolation Analysis in Polymer-Sorted
  (7,5) Single-Walled Carbon Nanotube Networks. \emph{Small} \textbf{2016},
  \emph{12}, 4211--4221\relax
\mciteBstWouldAddEndPuncttrue
\mciteSetBstMidEndSepPunct{\mcitedefaultmidpunct}
{\mcitedefaultendpunct}{\mcitedefaultseppunct}\relax
\EndOfBibitem
\bibitem[Hartleb \latin{et~al.}(2015)Hartleb, Sp{\"a}th, and
  Hertel]{Hartleb2015}
Hartleb,~H.; Sp{\"a}th,~F.; Hertel,~T. Evidence for Strong Electronic
  Correlations in the Spectra of Gate-Doped Single-Wall Carbon Nanotubes.
  \emph{ACS Nano} \textbf{2015}, \emph{9}, 10461--10470\relax
\mciteBstWouldAddEndPuncttrue
\mciteSetBstMidEndSepPunct{\mcitedefaultmidpunct}
{\mcitedefaultendpunct}{\mcitedefaultseppunct}\relax
\EndOfBibitem
\bibitem[Shi \latin{et~al.}(2018)Shi, Isakova, Abudulimu, {van den Berg}, Kwon,
  Meixner, Park, Zhang, Gierschner, and L{\"u}er]{Shi2018}
Shi,~J.; Isakova,~A.; Abudulimu,~A.; {van den Berg},~M.; Kwon,~O.~K.;
  Meixner,~A.~J.; Park,~S.~Y.; Zhang,~D.; Gierschner,~J.; L{\"u}er,~L.
  Designing High Performance All-Small-Molecule Solar Cells with Non-Fullerene
  Acceptors: Comprehensive Studies on Photoexcitation Dynamics and Charge
  Separation Kinetics. \emph{Energy Environ. Sci.} \textbf{2018}, \emph{11},
  211--220\relax
\mciteBstWouldAddEndPuncttrue
\mciteSetBstMidEndSepPunct{\mcitedefaultmidpunct}
{\mcitedefaultendpunct}{\mcitedefaultseppunct}\relax
\EndOfBibitem
\bibitem[Matsunaga \latin{et~al.}(2011)Matsunaga, Matsuda, and
  Kanemitsu]{Matsunaga2011}
Matsunaga,~R.; Matsuda,~K.; Kanemitsu,~Y. Observation of Charged Excitons in
  Hole-Doped Carbon Nanotubes Using Photoluminescence and Absorption
  Spectroscopy. \emph{Phys. Rev. Lett.} \textbf{2011}, \emph{106}, 037404\relax
\mciteBstWouldAddEndPuncttrue
\mciteSetBstMidEndSepPunct{\mcitedefaultmidpunct}
{\mcitedefaultendpunct}{\mcitedefaultseppunct}\relax
\EndOfBibitem
\bibitem[Nishihara \latin{et~al.}(2012)Nishihara, Yamada, and
  Kanemitsu]{Nishihara2012}
Nishihara,~T.; Yamada,~Y.; Kanemitsu,~Y. Dynamics of Exciton-Hole Recombination
  in Hole-Doped Single-Walled Carbon Nanotubes. \emph{Phys. Rev. B}
  \textbf{2012}, \emph{86}, 075449\relax
\mciteBstWouldAddEndPuncttrue
\mciteSetBstMidEndSepPunct{\mcitedefaultmidpunct}
{\mcitedefaultendpunct}{\mcitedefaultseppunct}\relax
\EndOfBibitem
\bibitem[Park \latin{et~al.}(2012)Park, Hirana, Mouri, Miyauchi, Nakashima, and
  Matsuda]{Park2012}
Park,~J.~S.; Hirana,~Y.; Mouri,~S.; Miyauchi,~Y.; Nakashima,~N.; Matsuda,~K.
  Observation of Negative and Positive Trions in the Electrochemically
  Carrier-Doped Single-Walled Carbon Nanotubes. \emph{J. Am. Chem. Soc.}
  \textbf{2012}, \emph{134}, 14461--14466\relax
\mciteBstWouldAddEndPuncttrue
\mciteSetBstMidEndSepPunct{\mcitedefaultmidpunct}
{\mcitedefaultendpunct}{\mcitedefaultseppunct}\relax
\EndOfBibitem
\bibitem[Mouri \latin{et~al.}(2013)Mouri, Miyauchi, Iwamura, and
  Matsuda]{Mouri2013}
Mouri,~S.; Miyauchi,~Y.; Iwamura,~M.; Matsuda,~K. Temperature Dependence of
  Photoluminescence Spectra in Hole-Doped Single-Walled Carbon Nanotubes:
  Implications of Trion Localization. \emph{Phys. Rev. B} \textbf{2013},
  \emph{87}\relax
\mciteBstWouldAddEndPuncttrue
\mciteSetBstMidEndSepPunct{\mcitedefaultmidpunct}
{\mcitedefaultendpunct}{\mcitedefaultseppunct}\relax
\EndOfBibitem
\bibitem[Gaulke \latin{et~al.}(2020)Gaulke, Janissek, Peyyety, Alamgir, Riaz,
  Dehm, Li, Lemmer, Flavel, Kappes, Hennrich, Wei, Chen, Pyatkov, and
  Krupke]{Gaulke2020}
Gaulke,~M.; Janissek,~A.; Peyyety,~N.~A.; Alamgir,~I.; Riaz,~A.; Dehm,~S.;
  Li,~H.; Lemmer,~U.; Flavel,~B.~S.; Kappes,~M.~M. \latin{et~al.}
  Low-Temperature Electroluminescence Excitation Mapping of Excitons and Trions
  in Short-Channel Monochiral Carbon Nanotube Devices. \emph{ACS Nano}
  \textbf{2020}, \emph{14}, 2709--2717\relax
\mciteBstWouldAddEndPuncttrue
\mciteSetBstMidEndSepPunct{\mcitedefaultmidpunct}
{\mcitedefaultendpunct}{\mcitedefaultseppunct}\relax
\EndOfBibitem
\bibitem[Eremin \latin{et~al.}(2019)Eremin, Obraztsov, Velikanov, Shubina, and
  Obraztsova]{Eremin2019}
Eremin,~T.~V.; Obraztsov,~P.~A.; Velikanov,~V.~A.; Shubina,~T.~V.;
  Obraztsova,~E.~D. Many-Particle Excitations in Non-Covalently Doped
  Single-Walled Carbon Nanotubes. \emph{Sci. Rep.} \textbf{2019}, \emph{9},
  14985\relax
\mciteBstWouldAddEndPuncttrue
\mciteSetBstMidEndSepPunct{\mcitedefaultmidpunct}
{\mcitedefaultendpunct}{\mcitedefaultseppunct}\relax
\EndOfBibitem
\bibitem[Huard \latin{et~al.}(2000)Huard, Cox, Saminadayar, Arnoult, and
  Tatarenko]{Huard2000}
Huard,~V.; Cox,~R.~T.; Saminadayar,~K.; Arnoult,~A.; Tatarenko,~S. Bound States
  in Optical Absorption of Semiconductor Quantum Wells Containing a
  Two-Dimensional Electron Gas. \emph{Phys. Rev. Lett.} \textbf{2000},
  \emph{84}, 187--190\relax
\mciteBstWouldAddEndPuncttrue
\mciteSetBstMidEndSepPunct{\mcitedefaultmidpunct}
{\mcitedefaultendpunct}{\mcitedefaultseppunct}\relax
\EndOfBibitem
\bibitem[Esser \latin{et~al.}(2001)Esser, Zimmermann, and Runge]{Esser2001}
Esser,~A.; Zimmermann,~R.; Runge,~E. Theory of Trion Spectra in Semiconductor
  Nanostructures. \emph{Phys. Status Solidi B} \textbf{2001}, \emph{227},
  317--330\relax
\mciteBstWouldAddEndPuncttrue
\mciteSetBstMidEndSepPunct{\mcitedefaultmidpunct}
{\mcitedefaultendpunct}{\mcitedefaultseppunct}\relax
\EndOfBibitem
\bibitem[Zorn \latin{et~al.}(2020)Zorn, Scuratti, Berger, Perinot, Heimfarth,
  Caironi, and Zaumseil]{Zorn2020}
Zorn,~N.~F.; Scuratti,~F.; Berger,~F.~J.; Perinot,~A.; Heimfarth,~D.;
  Caironi,~M.; Zaumseil,~J. Probing Mobile Charge Carriers in Semiconducting
  Carbon Nanotube Networks by Charge Modulation Spectroscopy. \emph{ACS Nano}
  \textbf{2020}, \emph{14}, 2412--2423\relax
\mciteBstWouldAddEndPuncttrue
\mciteSetBstMidEndSepPunct{\mcitedefaultmidpunct}
{\mcitedefaultendpunct}{\mcitedefaultseppunct}\relax
\EndOfBibitem
\bibitem[Liu and Xiao(2011)Liu, and Xiao]{Liu2011}
Liu,~T.; Xiao,~Z. Exact and Closed Form Solutions for the Quantum Yield,
  Exciton Diffusion Length, and Lifetime to Reveal the Universal Behaviors of
  the Photoluminescence of Defective Single-Walled Carbon Nanotubes. \emph{J.
  Phys. Chem. C} \textbf{2011}, \emph{115}, 16920--16927\relax
\mciteBstWouldAddEndPuncttrue
\mciteSetBstMidEndSepPunct{\mcitedefaultmidpunct}
{\mcitedefaultendpunct}{\mcitedefaultseppunct}\relax
\EndOfBibitem
\bibitem[Cognet \latin{et~al.}(2007)Cognet, Tsyboulski, Rocha, Doyle, Tour, and
  Weisman]{Cognet2007}
Cognet,~L.; Tsyboulski,~D.~A.; Rocha,~J.-D.~R.; Doyle,~C.~D.; Tour,~J.~M.;
  Weisman,~R.~B. Stepwise Quenching of Exciton Fluorescence in Carbon Nanotubes
  by Single-Molecule Reactions. \emph{Science} \textbf{2007}, \emph{316},
  1465--1468\relax
\mciteBstWouldAddEndPuncttrue
\mciteSetBstMidEndSepPunct{\mcitedefaultmidpunct}
{\mcitedefaultendpunct}{\mcitedefaultseppunct}\relax
\EndOfBibitem
\bibitem[Xie \latin{et~al.}(2012)Xie, Inaba, Sugiyama, and Homma]{Xie2012}
Xie,~J.; Inaba,~T.; Sugiyama,~R.; Homma,~Y. Intrinsic Diffusion Length of
  Excitons in Long Single-Walled Carbon Nanotubes from Photoluminescence
  Spectra. \emph{Phys. Rev. B} \textbf{2012}, \emph{85}, 085434\relax
\mciteBstWouldAddEndPuncttrue
\mciteSetBstMidEndSepPunct{\mcitedefaultmidpunct}
{\mcitedefaultendpunct}{\mcitedefaultseppunct}\relax
\EndOfBibitem
\bibitem[Georgi \latin{et~al.}(2009)Georgi, B{\"o}hmler, Qian, Novotny, and
  Hartschuh]{Georgi2009}
Georgi,~C.; B{\"o}hmler,~M.; Qian,~H.; Novotny,~L.; Hartschuh,~A. Probing
  Exciton Propagation and Quenching in Carbon Nanotubes with Near-Field Optical
  Microscopy. \emph{Phys. Status Solidi B} \textbf{2009}, \emph{246},
  2683--2688\relax
\mciteBstWouldAddEndPuncttrue
\mciteSetBstMidEndSepPunct{\mcitedefaultmidpunct}
{\mcitedefaultendpunct}{\mcitedefaultseppunct}\relax
\EndOfBibitem
\bibitem[Mann and Hertel(2016)Mann, and Hertel]{Mann2016}
Mann,~C.; Hertel,~T. 13 nm Exciton Size in (6,5) Single-Wall Carbon Nanotubes.
  \emph{J. Phys. Chem. Lett.} \textbf{2016}, \emph{7}, 2276--2280\relax
\mciteBstWouldAddEndPuncttrue
\mciteSetBstMidEndSepPunct{\mcitedefaultmidpunct}
{\mcitedefaultendpunct}{\mcitedefaultseppunct}\relax
\EndOfBibitem
\bibitem[Ma \latin{et~al.}(2005)Ma, Valkunas, Dexheimer, Bachilo, and
  Fleming]{Ma2005}
Ma,~Y.-Z.; Valkunas,~L.; Dexheimer,~S.~L.; Bachilo,~S.~M.; Fleming,~G.~R.
  Femtosecond Spectroscopy of Optical Excitations in Single-Walled Carbon
  Nanotubes: Evidence for Exciton-Exciton Annihilation. \emph{Phys. Rev. Lett.}
  \textbf{2005}, \emph{94}, 157402\relax
\mciteBstWouldAddEndPuncttrue
\mciteSetBstMidEndSepPunct{\mcitedefaultmidpunct}
{\mcitedefaultendpunct}{\mcitedefaultseppunct}\relax
\EndOfBibitem
\bibitem[Ueda \latin{et~al.}(2008)Ueda, Matsuda, Tayagaki, and
  Kanemitsu]{Ueda2008}
Ueda,~A.; Matsuda,~K.; Tayagaki,~T.; Kanemitsu,~Y. Carrier Multiplication in
  Carbon Nanotubes Studied by Femtosecond Pump-Probe Spectroscopy. \emph{Appl.
  Phys. Lett.} \textbf{2008}, \emph{92}, 233105\relax
\mciteBstWouldAddEndPuncttrue
\mciteSetBstMidEndSepPunct{\mcitedefaultmidpunct}
{\mcitedefaultendpunct}{\mcitedefaultseppunct}\relax
\EndOfBibitem
\bibitem[L{\"u}er \latin{et~al.}(2009)L{\"u}er, Hoseinkhani, Polli, Crochet,
  Hertel, and Lanzani]{Luer2009}
L{\"u}er,~L.; Hoseinkhani,~S.; Polli,~D.; Crochet,~J.; Hertel,~T.; Lanzani,~G.
  Size and Mobility of Excitons in (6, 5) Carbon Nanotubes. \emph{Nat. Phys.}
  \textbf{2009}, \emph{5}, 54--58\relax
\mciteBstWouldAddEndPuncttrue
\mciteSetBstMidEndSepPunct{\mcitedefaultmidpunct}
{\mcitedefaultendpunct}{\mcitedefaultseppunct}\relax
\EndOfBibitem
\bibitem[Kinder and Mele(2008)Kinder, and Mele]{Kinder2008}
Kinder,~J.~M.; Mele,~E.~J. Nonradiative Recombination of Excitons in Carbon
  Nanotubes Mediated by Free Charge Carriers. \emph{Phys. Rev. B}
  \textbf{2008}, \emph{78}, 155429\relax
\mciteBstWouldAddEndPuncttrue
\mciteSetBstMidEndSepPunct{\mcitedefaultmidpunct}
{\mcitedefaultendpunct}{\mcitedefaultseppunct}\relax
\EndOfBibitem
\bibitem[Perebeinos and Avouris(2008)Perebeinos, and Avouris]{Perebeinos2008}
Perebeinos,~V.; Avouris,~P. Phonon and Electronic Nonradiative Decay Mechanisms
  of Excitons in Carbon Nanotubes. \emph{Phys. Rev. Lett.} \textbf{2008},
  \emph{101}, 057401\relax
\mciteBstWouldAddEndPuncttrue
\mciteSetBstMidEndSepPunct{\mcitedefaultmidpunct}
{\mcitedefaultendpunct}{\mcitedefaultseppunct}\relax
\EndOfBibitem
\bibitem[Sau \latin{et~al.}(2013)Sau, Crochet, Doorn, and Cohen]{Sau2013}
Sau,~J.~D.; Crochet,~J.~J.; Doorn,~S.~K.; Cohen,~M.~L. Multiparticle Exciton
  Ionization in Shallow Doped Carbon Nanotubes. \emph{J. Phys. Chem. Lett.}
  \textbf{2013}, \emph{4}, 982--986\relax
\mciteBstWouldAddEndPuncttrue
\mciteSetBstMidEndSepPunct{\mcitedefaultmidpunct}
{\mcitedefaultendpunct}{\mcitedefaultseppunct}\relax
\EndOfBibitem
\bibitem[Berezhkovskii \latin{et~al.}(1990)Berezhkovskii, Makhnovskii, and
  Suris]{Berezhkovskii1990}
Berezhkovskii,~A.~M.; Makhnovskii,~Y.; Suris,~R.~A. Mean Square Displacement of
  a Brownian Particle with Traps. The One-Dimensional Case. \emph{Phys. Lett.
  A} \textbf{1990}, \emph{150}, 296--298\relax
\mciteBstWouldAddEndPuncttrue
\mciteSetBstMidEndSepPunct{\mcitedefaultmidpunct}
{\mcitedefaultendpunct}{\mcitedefaultseppunct}\relax
\EndOfBibitem
\end{mcitethebibliography}

\end{document}


\newpage
\section{Derivation of eq. 1 in the main text}

This section briefly describes how we arrive at eq. 1 by using scaling laws for diffusion-limited non-radiative decay. The derivation is thus based on the notion that PLQYs are governed by diffusive exciton transport to quenching sites which can be mobile or stationary. The derivation initially assumes that distances between each and every quenching site are identical. Variability of quenching site spacings is included in a second step by a weighted average over PL quantum yields, obtained for an Poissonian distribution of quenching site spacings.  


The change in the PL quantum yield $\Phi$ of the doped system with respect to the PL quantum yield $\Phi_0$ of the non-doped system is given by the respective rate constants $k$ and $k_0$ for diffusive transport to quenching sites in the doped and in the non-doped system, respectively.
\begin{equation}
	\frac{\Phi}{\Phi_0}=\frac{k_0}{k}
	\label{eq1}
\end{equation}
The reciprocal rate constant $k$ corresponds to a decay time $\tau$. For diffusion-limited processes this can also be related to the diffusion length $d$ via the diffusion coefficient $D$
\begin{equation}
    \frac{1}{k}=\tau = \frac{d^2}{2D}\text\,.
    \label{eq2}
\end{equation}
Next, we assume that the exciton diffusion coefficients in the intrinsic tube sections of the heterogeneously doped and non-doped system are identical and equation \ref{eq1} can be rewritten as
\begin{equation}
	\frac{\Phi}{\Phi_0}=\left[\frac{d_{\rm eff}}{d_0}\right]^2\text\,
	\label{eq3}
\end{equation}
where $d_0$ is the diffusion length needed for excitons to diffuse to quenching defects in non-doped SWNTs. This includes the effect of end-quenching as well as the contribution of other types of quenching sites found in nanotubes prepared by standard protocols.  Next, $d_{\rm eff}$ represents the effective diffusion length in doped SWNTs. The effective diffusion length is here determined by the sum of the reciprocal diffusion lengths, or equivalently, the joint concentration of quenching sites. This is given by 
\begin{equation}
	\frac{1}{d_{\rm eff}}=\frac{1}{d_0}+\frac{1}{d_d}\text\,
	\label{eq4}
\end{equation}
When calculating diffusion lengths from the intrinsic defect- and from the dopant- concentrations, $n_0$ and $n_d$, respectively, we need to account for the fact that the actual diffusion lengths are reduced somewhat by the spatial extent of defects as well as by the the spatial extent or size of the exciton. Accordingly we use
\begin{equation}
	d_0 = n_0^{-1}-\xi_X\text\,
	\label{eq5}
\end{equation}
and 
\begin{equation}
	d_d = n_d^{-1}-\xi_X-\xi_d\text\,
	\label{eq6}
\end{equation}
where $\xi_X$ and $\xi_d$ represent the exciton and defect sizes, respectively. The exciton size was previously determined to be on the order of 10 nm while the defect size was estimated to be about 4 nm \cite{Mann2016,Eckstein2019}. With this we obtain from eqs \ref{eq3},\ref{eq4},\ref{eq5} and \ref{eq6},
\begin{equation}
	\frac{\Phi}{\Phi_0}=\left[1+\frac{\xi_X-n_0^{-1}}{n_d^{-1}-\xi_d-\xi_X}\right]^{-2}
	\label{eq7}
\end{equation} 
Next, for comparison with experimental data we compute the concentration of quenching defects introduced by doping using the experimentally observed change of the normalized exciton oscillator strength $f$. This is done by using a simple phase space filling model where $f=1$ corresponds to the intrinsic, non-doped system. 
\begin{equation}
	f=1-\xi_d n_d\text\,
	\label{eq8}
\end{equation}
with $\xi_d$ again representing the size of the dopant-induced defects.

As mentioned above, this calculation implicitly assumes all distances between quenching defects to be identical. However, most doping schemes, such as the redox-chemical doping used in the experiments of this study, lead to a distribution of spacings between quenching sites which needs to be accounted for in the simulation. To do so we use a lattice model with randomly distributed doping sites. The exciton PLQY is then determined from an average over emission from intrinsic, i.e. undoped nanotube sections whose individual PLQYs are obtained from eq. \ref{eq7}.

Moreover, the intensity of emission from a specific tube segment is also weighed according to its length which represents a measure for the number of photons absorbed by that very segment.

The result of such averaging is then compared with the experimental data of Figure 6a) in the main text. To account for the fact that the diffusive model breaks down once the distance between two quenching sites becomes smaller than the size of the exciton we set the emission intensity of those segments to zero.

\section{Sample characterization}

\subsection{Sample composition}

\begin{figure}[htbp]
	\centering
		\includegraphics[width=16.75 cm]{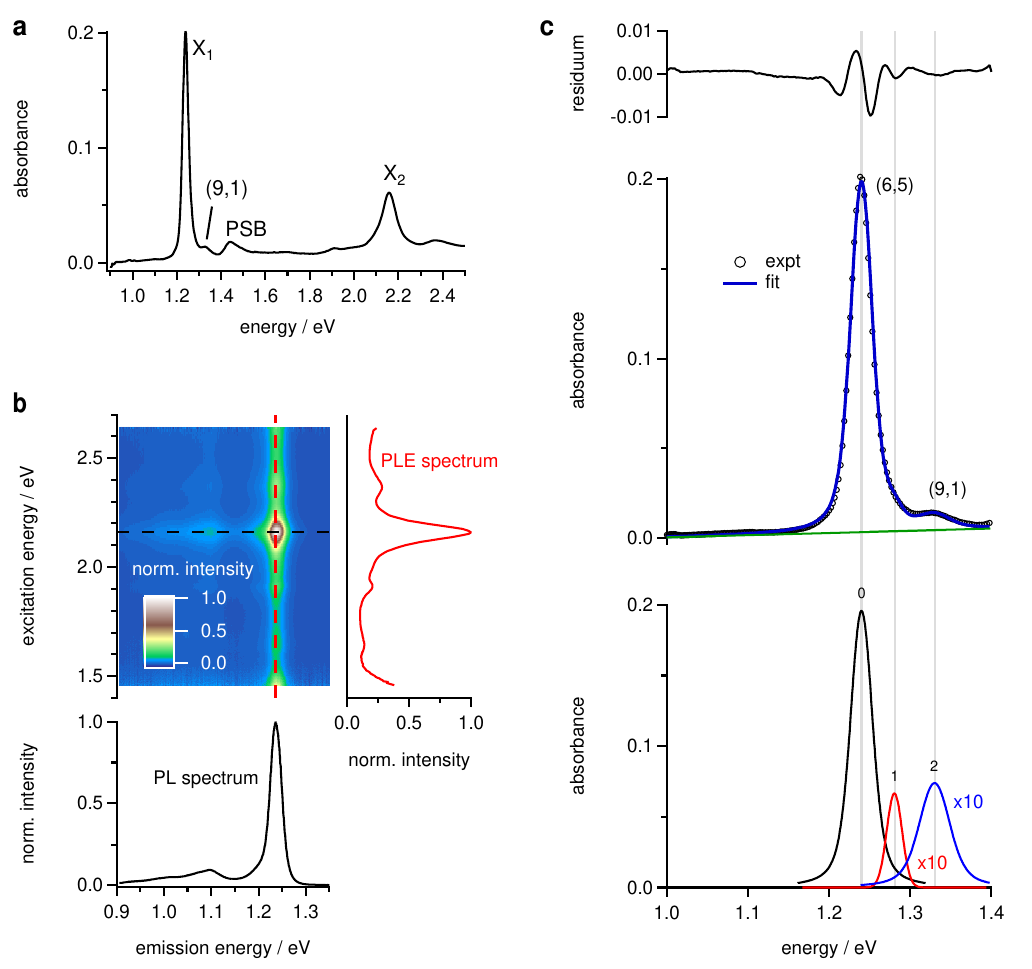}
		\caption{{\bf a)} Absorption and {\bf b)} photoluminescence excitation (PLE) spectrum of the as-prepared (6,5)-SWNT suspension. PSB = phonon sideband. {\bf c)} Multipeak fit used for the determination of (6,5)-purity.}
		\label{SI_fig1}
\end{figure}

Figure~\ref{SI_fig1}a) shows the absorption spectrum of the intrinsic nanotube suspension highly enriched in (6,5) chirality. Besides the $X_1$ exciton absorption band of the (9,1) minority species at 1.33\,eV, all peaks can be attributed to the (6,5) chirality as confirmed by photoluminescence excitation (PLE) spectroscopy (see Figure~\ref{SI_fig1}b)). The abundance of the different chiralities was calculated from the respective $X_1$ exciton bands of the multi-Voigt fit shown in Figure~\ref{SI_fig1}c), for simplicity assuming identical exciton oscillator strengths for both identified tube-types. The resulting percentages are shown in table~\ref{SI_tab1}.

\begin{table}[htbp]
    \centering
    \begin{tabular}{c|c|c}
       peak number  &  chirality & relative abundance / \% \\
         \hline 
         0 & (6,5) & 93 \\
         1 & unknown & 2 \\
         2 & (9,1) & 5
    \end{tabular}
    \caption{Relative abundance of different SWNT chiralities.}
    \label{SI_tab1}
\end{table}

\subsection{Charge carrier concentrations}

Table~\ref{SI_tab2} shows the correlation of carrier concentrations $n$ and dopant concentrations $c_{\rm AuCl_3}$. Carrier concentrations were obtained using the confinement analysis as reported by Eckstein et al.~\cite{Eckstein2019}.

\begin{table}[htbp]
    \centering
    \begin{tabular}{c|c||c|c}
       $c_{\rm AuCl_3} / \rm \mu g\,ml^{-1}$  &  $n\rm / nm^{-1}$  & $c_{\rm AuCl_3} / \rm \mu g\,ml^{-1}$  &  $n\rm / nm^{-1}$ \\
         \hline 
         0.00 &  & 1.2 &  0.039 \\
         0.03 &  & 1.5 &  0.064\\
         0.05 &  & 2.0 & 0.091\\
         0.07 &  & 2.5 &  0.11 \\
         0.10 &   0.008 & 3.0 & 0.12 \\
         0.30 &  0.013 & 4.0 & 0.13 \\
         0.50 & 0.017 & 6.0 &  0.15 \\
         0.70 &  0.021 & 8.0 &  0.16\\
         0.90 & 0.026 & 15 &  0.17\\
         1.0 &  0.030 &    & \\
    \end{tabular}
    \caption{Charge carrier concentrations corresponding to the absorption and PL spectra shown in Figures~1 and 3a.}
    \label{SI_tab2}
\end{table}

\section{Density of excited states in pump probe experiments}

In order to exclusively study the doping-induced changes of exciton and trion dynamics, the excitation density $\rho$ should be considerably below the density of quenching defects $n$. Otherwise, bimolecular de-excitation processes such as exciton-exciton annihilation (EEA) can become relevant.
The excitation density is shown in table~\ref{SI_tab3} and was calculated via

\begin{equation}
    \rho = F\,\sigma\,c,
\end{equation}

where $F$ is the pump pulse fluences at the sample position, $\sigma$ is the absorption cross section per carbon atom and $c=88\,\rm nm^{-1}$ is the number of carbon atoms per nanometer of (6,5)-SWNT tube length. The absorption cross section at the excitation energy is an estimate based on the comparison of absorption spectra with published data by Schöppler~\textit{et al.}\cite{Schoeppler2011}. Here, we use an absorption cross section $\sigma = 6\times 10^{-18}\,\rm cm^2$ at the $X_2$ peak of the intrinsic sample. The tabulated cross sections at the exciton and trion energies in the doped suspension are referenced to this value according to their relative absorbance (see also Figure~\ref{SI_fig3}a). 

\begin{table}[htbp]
    \centering
    \begin{tabular}{c|c|c|c|c}
       shown in  &  photon energy / eV & $F / 10^{13}\,\rm cm^{-2}$ &  $\sigma / 10^{-18}\,\rm cm^2$ & $\rho / \rm nm^{-1}$ \\
         \hline 
         Fig. 7a, 7b, 9 & 2.15 & 0.29 & 6 & 1/650 \\
         Fig. 8 & 1.24 & 3.9 & 4 & 1/70 \\
         Fig. 7a, 8 & 1.06 & 2.1 & 2 & 1/270 
    \end{tabular}
    \caption{Experimental conditions for the pump probe experiments shown in Figures~7 and 8.}
    \label{SI_tab3}
\end{table}

\section{Full 2D transient absorption spectra}

\begin{figure}[htbp]
	\centering
		\includegraphics[width=17.9 cm]{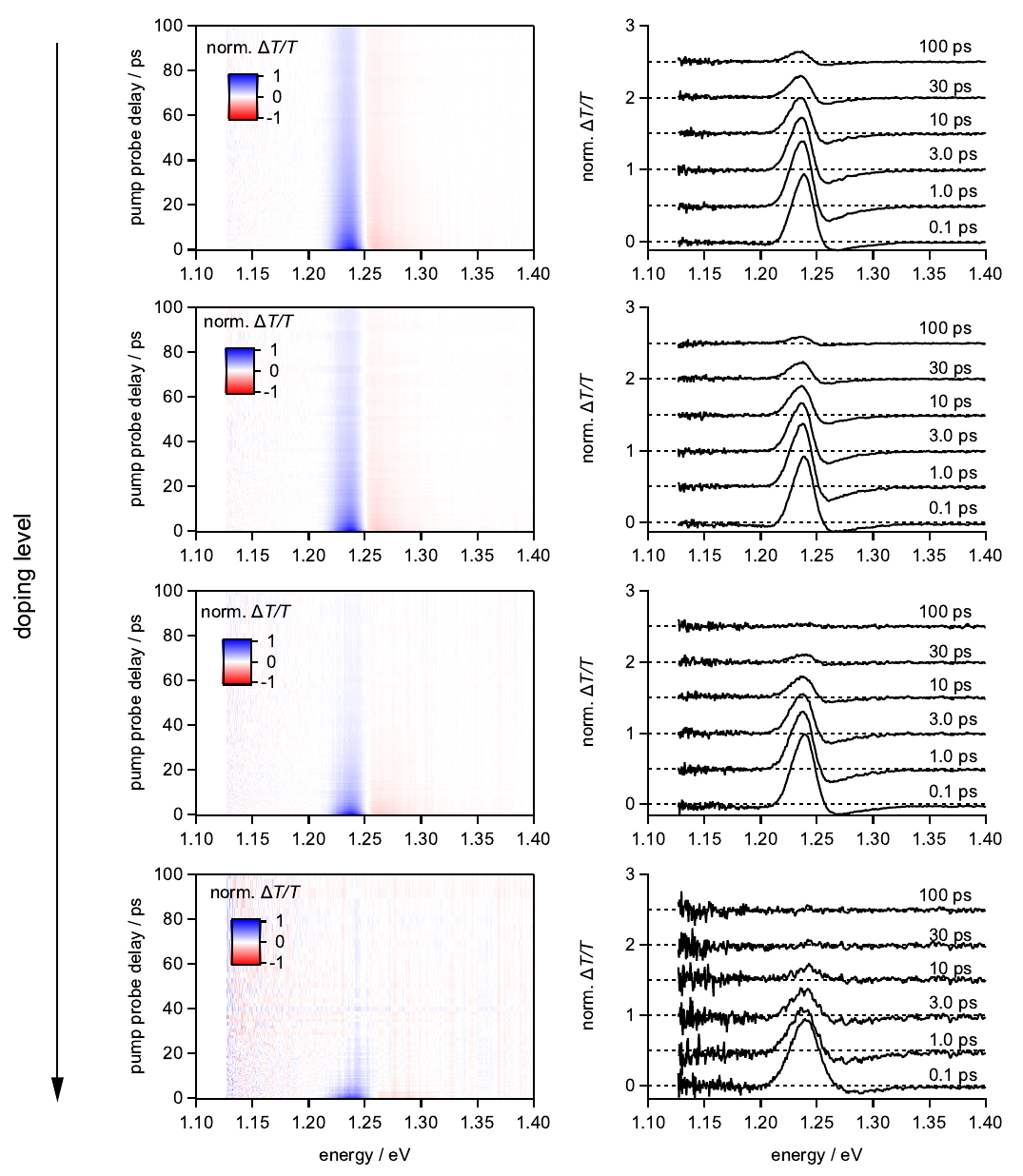}
		\caption{Transient absorption spectra used for the evaluation of exciton dynamics.}
		\label{SI_fig2}
\end{figure}

Figure~\ref{SI_fig2} shows 2D transient absorption spectra (left) and single transient spectra at selected delay times (right) used for the evaluation of exciton dynamics in Figures~7a and 7b. Note the low energy cutoff at $\approx 1.13\,\rm eV$ due to the usage of a silicon CCD camera detector.

\begin{figure}[htbp]
	\centering
		\includegraphics[width=14.0 cm]{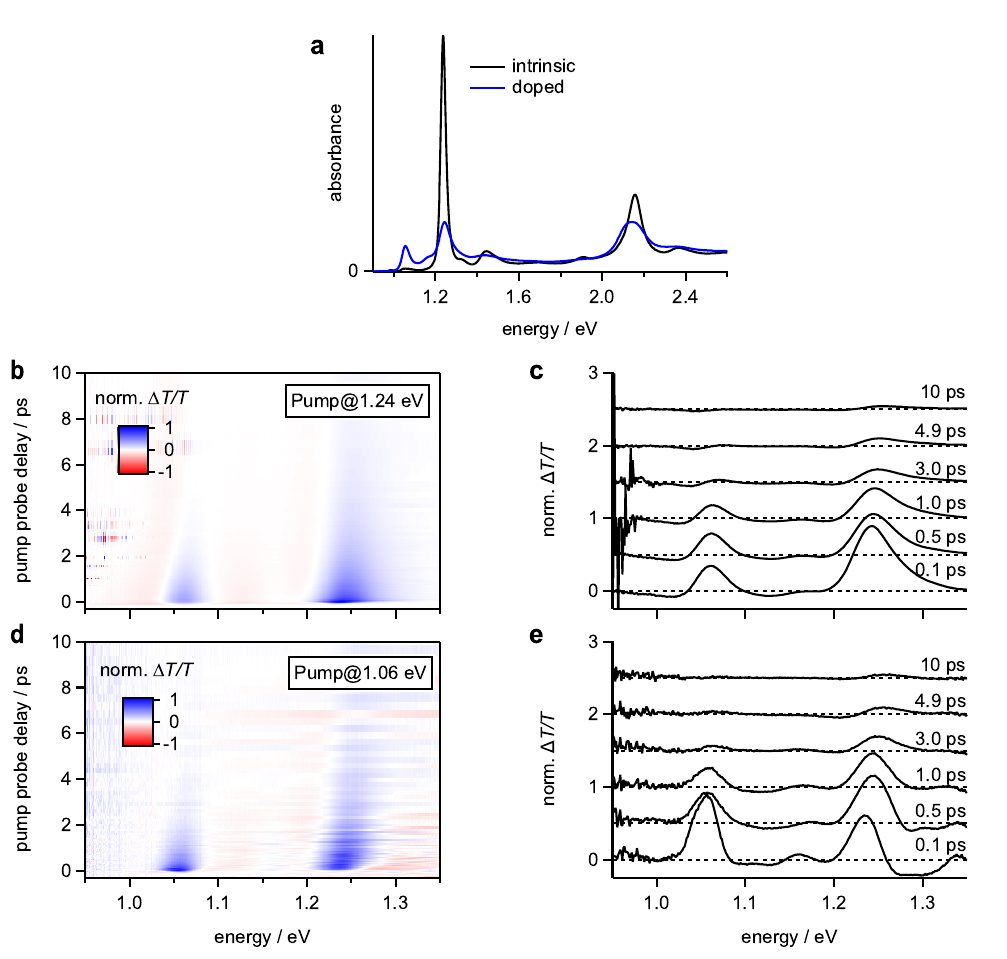}
		\caption{Linear and transient absorption spectra used for the evaluation of trion dynamics.}
		\label{SI_fig3}
\end{figure}

Figure~\ref{SI_fig3} shows the absorption spectra of the nanotube suspension, 2D transient absorption spectra (left) and transient spectra at selected delay times (right) used for the evaluation of trion dynamics in Figures~7a and 8.

\section{Global analysis of trion photobleach and adjacent photoabsorption signal}

\begin{figure}[htbp]
	\centering
		\includegraphics[width=16.7 cm]{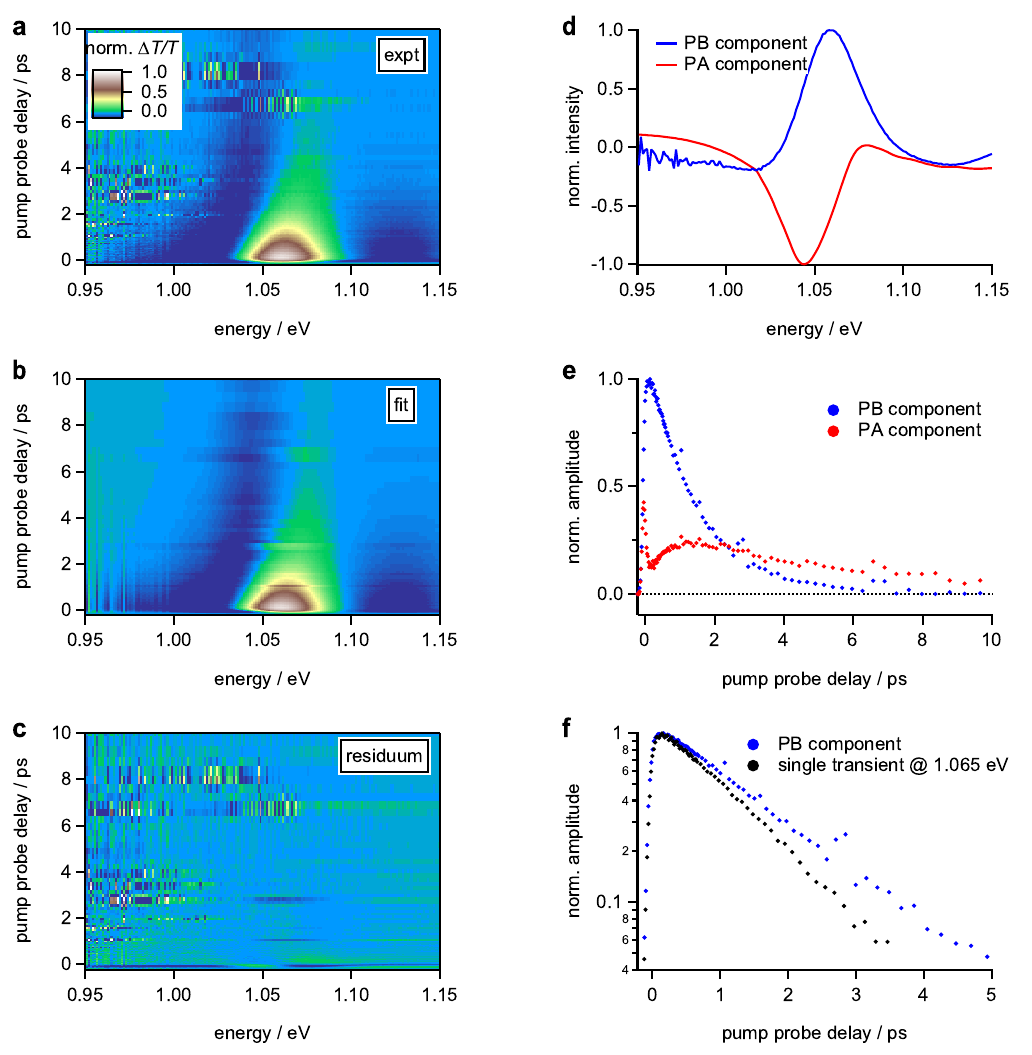}
		\caption{{\bf a)} Experimental 2D transient absorption spectrum, {\bf b)} results of a global analysis and {\bf c)} the residuum. {\bf d)} The two normalized spectral fit components and {\bf e)} their normalized weight as a function of pump-probe delay. {\bf f)} Comparison between the dynamics of the PB component and a single transient at the trion PB maximum.}
		\label{SI_fig4}
\end{figure}

The trion photobleach (PB) signal is spectrally overlapping with a slightly redshifted photoabsorption (PA) signal. 
The relative weight of PA with respect to PB rises with doping level but decreases when resonantly pumping at 1.06\,eV compared to other excitation conditions, such as resonantly pumping at the exciton resonance (see Figure~\ref{SI_fig3}b-e). 
Actually, for doping levels where both the $X_1$ and $X_1^+$ resonance are fully bleached in the absorption spectrum, we do not observe a trion PB feature. In this case, the PA signal is dominating the transient spectral response.
Figure~\ref{SI_fig4} illustrates the spectral deconvolution for the experiment shown in Figure~8) under resonant excitation of the $X_1$ exciton. This deconvolution is achieved by a global analysis using two spectral components, called the PA and PB component. Here, the PA component was approximated using averaged transient spectra at later delay times (9-11\,ps) where the positive photobleach signal has already nearly vanished. Additionally, we smoothed the PA component due to a poor signal-to-noise ratio for these low-intensity spectra. For the fit, we fixed the PA component, whereas the spectral PB component as well as the time-dependent amplitudes of both PA and PB were free fit parameters. The trion dynamics is here associated with the dynamics of the PB component. Note the spike in the dynamics of the PA component around time zero. This is due to ignoring the coherent artifact in our global analysis, which is also evident from the considerable residuum at temporal pump-probe overlap.
As can be seen from Figure~\ref{SI_fig4}f the decay dynamics of the PB component are only slightly different from a single transient at 1.065\,eV, where the trion PB maximum is located.

